\documentstyle[epsfig]{aipproc}

\begin{document}
\title{Neutrino Physics}
\author{R. D. Peccei}
\address{Department of Physics and Astronomy, UCLA, Los Angeles, CA 
90095-1547}
\maketitle

\begin{abstract}
These lectures describe some aspects of the physics of massive neutrinos. After a brief introduction of neutrinos in the Standard Model, I discuss possible patterns for their masses. In particular, I show how the presence of a large Majorana mass term for the right-handed neutrinos can engender tiny neutrino masses for the observed neutrinos. If neutrinos have mass, different flavors of neutrinos can oscillate into one another. To analyze this phenomena, I develop the relevant formalism for neutrino oscillations, both in vacuum and in matter. After reviewing the existing (negative) evidence for neutrino masses coming from direct searches, I discuss evidence for, and hints of, neutrino oscillations in the atmosphere, the sun, and at accelerators.  Some of the theoretical implications of these results are emphasized.  I close these lectures by briefly outlining future experiments which will shed further light on atmospheric, accelerator and solar neutrino oscillations. A pedagogical discussion of Dirac and Majorana masses is contained in an appendix.
\end{abstract}

\section{Neutrinos in the Standard Model}

Neutrinos play a special role in the $SU(2)\times U(1)$
electroweak theory.  While the left-handed neutrinos are
part of $SU(2)$ doublets
\begin{equation}
L_i = \left(
\begin{array}{c}
\nu_{\ell_i} \\ \ell_i
\end{array} \right)_{\rm L}~, ~~~~~~ \ell_i = \{e,\mu,\tau\}~,
\end{equation}
the right-handed neutrinos are $SU(2)$ singlets.  Since the
electromagnetic charge and the $U(1)$ hypercharge differ by
the value of the third component of weak isospin
\begin{equation}
Q = T_3 + Y~,
\end{equation}
one sees that the right-handed neutrinos $(\nu_{\ell_i})_{\rm R}$
carry {\bf no} $SU(2) \times U(1)$ quantum numbers.  The
above has two important experimental implications:
\begin{description}
\item{i)} The neutrinos seen experimentally are those produced
by the weak interactions.  These neutrinos are purely
left-handed.
\item{ii)} Because one cannot infer the existence of right-handed
neutrinos from weak processes, the presence of $(\nu_{\ell_i})_{\rm R}$ can only be seen indirectly, most likely
through the existence of neutrino masses.  However, 
it should be noted that neutrino
masses do not necessarily imply the existence of right-handed
neutrinos.
\end{description}

Before discussing issues connected with neutrino masses, it is
useful to summarize what we know about the left-handed neutrinos
from their weak interactions.  In the electroweak theory  
these neutrinos couple to the
$Z$-boson in a universal fashion
\begin{equation}
{\cal{L}}_{Z\nu\bar\nu} = \frac{e}{2\cos\theta_W\sin\theta_W}
Z^\mu[J_\mu^{\rm NC}]_\nu~,
\end{equation}
where
\begin{equation}
[J_\mu^{\rm NC}]_\nu = \sum_i (\bar\nu_{\ell_i})_{\rm L}
\gamma_\mu (\nu_{\ell_i})_{\rm L}~.
\end{equation}
Precision studies of the $Z$-line shape allow the determination
of the number of neutrino species $i$, since as this number
increases so does the $Z$ total width.  Each neutrino type
(provided its mass $m_{\nu_{i}} \ll \frac{M_Z}{2}$)
contributes the same amount to the $Z$-width\cite{LEPEW}
\begin{equation}
\Gamma(Z\to\nu_{\ell_i}\bar\nu_{\ell_i}) =
\frac{\sqrt{2}G_FM_Z^3}{24\pi} \rho~.
\end{equation}
Here $G_F$ is the Fermi constant as determined in $\mu$-decay\cite{PDG} and $\rho$ is related to the axial coupling
of the charged leptons to the $Z$-boson: $\rho = (2g_A^\ell)^2$.
Using\cite{LEPEW}
\begin{equation}
g_A^\ell = -0.50102 \pm 0.00030~,
\end{equation}
which is the average of the results obtained by the four LEP
collaborations and SLD, one has numerically
\begin{equation}
\Gamma_\nu \equiv \Gamma (Z\to\nu_{\ell_i}\bar\nu_{\ell_i}) = (167.06 \pm 0.22) ~{\rm MeV}~.
\end{equation}
Using the above, the number of different neutrino species, $N_\nu$, can 
then be
derived from the precision measurements of the $Z$ total
width and of its partial width into hadrons and leptons using
the (obvious) equation
\begin{equation}
\Gamma_{\rm tot} = \Gamma_{\rm had} + 3\Gamma_{\rm lept} +
N_\nu\Gamma_\nu~.
\end{equation}

From a fit of the $Z$-line shape for $e^+e^-\to\mu^+\mu^-$ and
$e^+e^-\to$ hadrons at LEP one can extract very accurate
values for $\Gamma_{\rm tot},~\Gamma_{\rm lept}$ and
$\Gamma_{\rm had}$:
\begin{eqnarray}
\Gamma_{\rm tot} &=& (2.4939 \pm 0.0024)~{\rm GeV}\nonumber \\
\Gamma_{\rm lept} &=& (83.90 \pm 0.1)~{\rm MeV} \nonumber \\
\Gamma_{\rm had} &=& (1.7423 \pm 0.0023)~{\rm GeV}~.
\end{eqnarray}
Whence it follows that the, so-called, invisible width is
\begin{equation}
\Gamma_{{\rm inv}} = N_\nu\Gamma_\nu = (499.9 \pm 3.4)~{\rm MeV}
\end{equation}
and thus one deduces for $N_\nu$---the number of neutrino
species:
\begin{equation}
N_\nu = 2.992 \pm 0.020~.
\end{equation}
This result strongly supports the notion, expressed in Eq. (1), that there
are only 3 generations of leptons.

It turns out that one can get a more accurate value for
$N_\nu$ by using other information derivable from the $Z$-line
shape.  The cross-section for $e^+e^-\to$ hadrons can be
expressed in terms of three factors:\cite{LEPZ} a peak cross-section
\begin{equation}
\sigma_o = \frac{12\pi\Gamma_{\rm lept}\Gamma_{\rm had}}
{\Gamma_{\rm tot}^2 M_Z^2}~,
\end{equation}
a Breit-Wigner factor
\begin{equation}
{\rm BW}(s) = \frac{s~\Gamma_{\rm tot}^2}
{(s-M^2_Z)^2+s^2\Gamma_{\rm tot}^2/M_Z^2}~,
\end{equation}
and a computable initial state bremsstrahlung correction
$(1-\delta_{\rm QED}(s))$, with
\begin{equation}
\sigma_{\rm had} = \sigma_o~{\rm BW}(s)(1-\delta_{\rm QED}(s))~.
\end{equation}
The value of $\sigma_o$ extracted from an analysis of the
cross-section for $e^+e^-\to$ hadrons at LEP\cite{LEPEW}
\begin{equation}
\sigma_o = (41.491 \pm 0.058)~{\rm nb}
\end{equation}
can be combined with the LEP results for the ratio of hadronic
to leptonic partial widths of the $Z$
\begin{equation}
R_\ell = \frac{\Gamma_{\rm had}}{\Gamma_{\rm lept}} =
20.765 \pm 0.026
\end{equation}
to deduce, with a little bit of theoretical input, a value for
$N_\nu$.  This value, as we will see, is slightly more
accurate than that given in Eq. (11).

Using Eqs. (8), (10), (12), and (16), one can write
\begin{equation}
N_\nu = \frac{\Gamma_{\rm inv}}{\Gamma_\nu} =
\frac{\Gamma_{\rm lept}}{\Gamma_\nu}
\left\{\frac{\Gamma_{\rm tot}}{\Gamma_{\rm lept}} -
\frac{\Gamma_{\rm had}}{\Gamma_{\rm lept}} - 3\right\} =
\frac{\Gamma_{\rm lept}}{\Gamma_\nu}
\left\{\sqrt{\frac{12\pi R_\ell}{\sigma_o M^2_Z}} -
R_\ell - 3\right\}~.
\end{equation}
In the Standard Model, the ratio $\Gamma_{\rm lept}/\Gamma_\nu$
is very accurately known:
\begin{equation}
\left.\frac{\Gamma_\nu}{\Gamma_{\rm lept}}\right|_{\rm SM} =
1.991 \pm 0.001~.
\end{equation}
Using this value in Eq. (17), along with the experimentally
determined $Z$ mass $M_Z = (91.1867 \pm 0.0021)$ GeV and the
values of $\sigma_o$ and $R_\ell$ measured at LEP, gives
\begin{equation}
N_\nu = 2.994 \pm 0.011~; ~~~~
\Gamma_{{\rm inv}} = (500.1 \pm 1.9)~{\rm MeV}~.
\end{equation}
These values are consistent with those in Eqs. (10) and (11), but are
about a factor of two more accurate.

There is an analogous equation to Eq. (3) describing the
coupling of the
$W^\pm$ boson to the leptonic charged currents.  Again
only left-handed neutrinos are involved.  One has
\begin{equation}
{\cal{L}}_{W\ell\nu_\ell} = \frac{e}{\sqrt{2}\sin\theta_W}
\{W^\mu_+ J_{\mu -}^{\rm lept} + W^\mu_- J_{\mu +}^{\rm lept}
\}~,
\end{equation}
where
\begin{equation}
J_{\mu -}^{\rm lept} = (J_{\mu +}^{\rm lept})^\dagger =
\sum_i \bar\ell_{i{\rm L}}\gamma_\mu\nu_{\ell_i{\rm L}}~.
\end{equation}
The states that appear in Eq. (21) in general are not
mass eigenstates, since mass generation can mix leptons of the
same charge among each other.  Nevertheless, one can always
diagonalize the charged lepton mass matrix by a by-unitary
transformation of the left- and right-handed charged lepton fields:
\begin{equation}
\ell_{\rm L} = U^\ell\tilde\ell_{\rm L}~; ~~~~~
\ell_{\rm R} = V^\ell\tilde\ell_{\rm R}~.
\end{equation}
After this transformation, the charged currents in Eq. (21) read
\begin{equation}
J_{\mu -}^{\rm lept} = (J_{\mu +}^{\rm lept})^\dagger = \sum_{ij}
{\overline{\tilde\ell_{i{\rm L}}}}
\gamma_\mu (U^\ell)^\dagger_{ij}
\nu_{\ell_j{\rm L}} = \sum_i {\overline{\tilde\ell_{i{\rm L}}}}\gamma_\mu
\tilde \nu_{\ell_i{\rm L}}~,
\end{equation}
where
\begin{equation}
\tilde \nu_{\ell {\rm L}} = (U^\ell)^\dagger \nu_{\ell L}~.
\end{equation}

Note that because $U^\ell$ is unitary, the neutral current
$[J_\mu^{\rm NC}]_\nu$ is the same whether it is expressed in
terms of $\nu_{\ell{\rm L}}$ or $\tilde\nu_{\ell{\rm L}}$.
Conventionally, the states $\tilde\nu_{\ell_i{\rm L}}$ are
called {\bf weak interaction eigenstates}, since they are
produced in the decay of a $W^+$ boson in association with a physically
charged lepton $\tilde\ell_i$ of definite mass.  These states, of
course, are also pair produced by the $Z$-boson.  For ease of 
notation, in what follows I will drop the tilde on both
$\tilde\ell_{i{\rm L}}$ and $\tilde\nu_{\ell_{i{\rm L}}}$ with the
understanding that 
the states now called $\nu_{\ell_i{\rm L}}$ are those produced by the weak interactions---they 
are the weak interaction eigenstates.  Similarly, the charged leptons $\ell_i$ are
the states associated with a diagonal mass matrix
\begin{equation}
M_\ell = \left(
\begin{array}{ccc}
m_e   &  & \\
& m_\mu &  \\
&  &  m_\tau
\end{array} \right)~.
\end{equation}

\section{Patterns of Neutrino Masses}

With these preliminaries underway, I want now to examine in a bit of detail
the possible patterns of neutrino masses.  To do so, it is useful to first
review how fermion masses originate in field theory.  The mass term with which
everybody is acquainted with is one involving a fermion $\psi$ and its conjugate
$\bar\psi = \psi^\dagger\gamma^0$:
\begin{equation}
{\cal{L}}_{\rm mass} = -m\bar\psi\psi = -m(\overline{\psi_{\rm L}}\psi_{\rm R} +
\overline{\psi_{\rm R}}\psi_{\rm L})~,
\end{equation}
with $\psi_{\rm L}$, $\psi_{\rm R}$ being the usual projections
\begin{equation}
\psi_{\rm L} = \frac{1}{2} (1-\gamma_5)\psi~; ~~~~
\psi_{\rm R} = \frac{1}{2} (1+\gamma_5)\psi~.
\end{equation}
This term, obviously, conserves fermion number
\begin{equation}
\psi\to e^{i\alpha}\psi~; ~~~~~ \bar\psi\to e^{-i\alpha}\bar\psi
\end{equation}
and gives equal mass for particles and antiparticles
\begin{equation}
m_{\bar\psi} = m_\psi = m~.
\end{equation}
For particles carrying any $U(1)$ quantum number, like electromagnetic charge,
it is clear that ${\cal{L}}_{\rm mass}$ is the {\bf only} possible mass term,
since to preserve these $U(1)$ quantum numbers one needs always to have particle-antiparticle interactions.

Neutrinos, however, provide an interesting exception.  Because  
neutrinos do not have
electromagnetic charge, it is possible to contemplate other types of mass terms
for them besides the particle-antiparticle term given in Eq. (26).  These 
other neutrino mass terms,
contain two neutrino (or two antineutrino) fields.  Hence they violate fermion
number (and in some cases $SU(2)\times U(1)$), but otherwise are 
allowed by Lorentz invariance.

As we discuss in more detail in Appendix A,
one can write three different types of mass
terms for neutrinos:
\begin{eqnarray}
{\cal{L}}_{\rm mass}^\nu = &-&[\overline{\nu_{\rm R}}m_D\nu_{\rm L} +
\overline{\nu_{\rm L}}m_D^\dagger\nu_{\rm R}] - \frac{1}{2}
[\overline{\nu_{\rm R}}\tilde C m_S\overline{\nu_{\rm R}}^T + \nu^T_{\rm R}
\tilde Cm^\dagger_S\nu_{\rm R}] \nonumber \\
&-& \frac{1}{2}[\nu_{\rm L}^T\tilde Cm_T\nu_{\rm L} +
\overline{\nu_{\rm L}}\tilde C m_T^\dagger\overline{\nu_{\rm L}}^T]~.
\end{eqnarray}
Here the mass matrices $m_D,m_S,m_T$ are Lorentz scalars.
However, their presence is only possible as a result of different symmetry
breakdowns.  Specifically, $m_D$ conserves fermion number, but violate
$SU(2)\times U(1)$ since it does not transform as an $SU(2)$ doublet.  This
fermion number conserving mass is often called a Dirac mass. Thus, in a happy
confluence of notation, $m_D$ can stand both for a Dirac mass and a doublet
mass.  Both $m_S$ and $m_T$ violate fermion number by two units and are known
as Majorana masses.  Because $m_S$ couples $\nu_{\rm R}$ with itself,
clearly it is an $SU(2)\times U(1)$ invariant.  This is not the case for
$m_T$, which violates $SU(2)\times U(1)$ because it does not transform as
an $SU(2)$ triplet.

The matrix $\tilde C$ which enters in the Majorana mass terms in Eq. (30)
is there to preserve Lorentz invariance.  Appendix A contains a detailed
discussion of this point, along with a pedagogical review of how one
constructs 4-spinors starting from 2-dimensional Weyl spinors.  I note here
only that $\tilde C$ is {\bf not} to be confused with the matrix $C$ connected
with how Dirac fields transform under charge conjugation.\cite{PecceiS}\footnote{Unfortunately, the distinction between
$C$ and $\tilde C$ is often blurred in the literature.}  Under the charge
conjugation operator $U(C)$ a Dirac field $\psi$ is transformed into its
Hermitian conjugate $\psi^\dagger$:
\begin{equation}
U(C) \psi U(C)^{-1} = C\psi^\dagger(x)~.
\end{equation}
The matrix $C$ is necessary to insure the invariance of the Dirac equation
under charge conjugation and obeys the restriction
\begin{equation}
C\gamma_\mu^* C^{-1} = -\gamma_\mu~.
\end{equation}
In general, $C$ depends on the $\gamma$-matrix basis used.  In the Majorana basis,
where the $\gamma$-matrices are purely imaginary, then $C=1$.  At any rate,
the matrix $\tilde C$ appearing in Eq. (30) is related to $C$ by \cite{BD}
\begin{equation}
\tilde C = C\gamma^{oT}~.
\end{equation}
It is easy to check that instead of (32) $\tilde C$ obeys
\begin{equation}
\tilde C\gamma_\mu^T \tilde C^{-1} = -\gamma_\mu~.
\end{equation}

The reason the matrix $\tilde C$ appears in Eq. (30) is because it relates the,
so-called, charge conjugate field $\psi^c$ to $\bar\psi$ rather than to
$\psi^\dagger$.  In view of the way the charge conjugation operator acts on the 
fermion field $\psi$ (see Eq. (31)), it is natural to define the charge
conjugate field $\psi^c$ as
\begin{equation}
\psi^c(x) = C\psi^\dagger(x)~.
\end{equation}
Now $\bar\psi = \psi^\dagger\gamma^0$, so one also has that
\begin{equation}
\psi^c(x) = C\gamma^{0T}\bar\psi^T(x) = \tilde C\bar\psi^T(x)~.
\end{equation}
So $\tilde C$, indeed, serves to relate $\bar\psi$ to $\psi^c$.

\section{The See-Saw Mechanism}

Eq. (30) displays the most general neutrino mass term, involving
three distinct mass matrices $m_D$, $m_S$ and $m_T$.  If there are only
three flavors of neutrinos these are $3\times 3$ matrices.  
For the moment this is what we shall
assume, but we shall return to this point later
on.  

One can write Eq. (30) in a more symmetrical way by replacing the
transposed fields in this equation by the charge conjugate field.
Recall that [cf. Eq. (35)]
\begin{equation}
\psi^c = \tilde C\bar\psi^T~; ~~~~
{\overline{\psi^c}} = \psi^T\tilde C~.
\end{equation}
Hence, for example, one can write\footnote{The minus sign
in the second term below comes from Fermi
statistics.}
\begin{equation}
\overline{\nu_{\rm R}}\nu_{\rm L} = -\nu^T_{\rm L}\overline{\nu_{\rm R}}^T =
\nu^T_{\rm L}\tilde C\tilde C\overline
{\nu_{\rm R}}^T = {\overline{\nu_{\rm L}^c}} \nu_{\rm R}^c~.
\end{equation}
Thus
\begin{equation}
\overline{\nu_{\rm R}}\nu_{\rm L} = \frac{1}{2}[\overline{\nu_{\rm R}}\nu_{\rm L} +
{\overline{\nu_{\rm L}^c}}\nu_{\rm R}^c]~.
\end{equation}
Using these equations, ${\cal{L}}^\nu_{\rm mass}$ can be written in the
following compact way:
\begin{equation}
{\cal{L}}^\nu_{\rm mass} = -\frac{1}{2}
\left[({\overline{\nu_{\rm L}^c}}~{\overline{\nu_{\rm R}}}) \left(
\begin{array}{cc}
m_T & m_D^T \\ m_D & m_S 
\end{array} \right)\left(
\begin{array}{c}
\nu_{\rm L} \\ \nu_R^c
\end{array} \right) \right]
+ {\rm h.c.}
\end{equation}

For 3 generations of neutrinos, the six mass eigenstates $m_i$ are the
eigenvalues of the $6\times 6$ matrix
\begin{equation}
M = \left(
\begin{array}{cc}
m_T & m_D^T \\ m_D & m_S 
\end{array} \right)~.
\end{equation}
Because $M$ is not necessarily Hermitian, its diagonalization necessitates
a bi-unitary transformation
\begin{equation}
U^\dagger_{\rm R} MU_{\rm L} = M_{\rm diag}~,
\end{equation}
where $U_{\rm L}$ and $U_{\rm R}$ are $6\times 6$ unitary matrices.  This
diagonalization is accomplished by a basis change on the original neutrino
fields
\begin{equation}
\psi_{\rm L} = \left(
\begin{array}{c}
\nu_{\rm L} \\ \nu_{\rm R}^c
\end{array} \right)~; ~~~~
\psi_{\rm R} = \left(
\begin{array}{c}
\nu_{\rm L}^c \\ \nu_{\rm R}
\end{array} \right)
\end{equation}
to a new set of fields $\eta_{\rm L}$ and $\eta_{\rm R}$ defined by the equations:
\begin{equation}
\psi_{\rm L} = U_{\rm L}\eta_{\rm L}~; ~~~~
\psi_{\rm R} = U_{\rm R}\eta_{\rm R}~.
\end{equation}

It is useful to consider the simple, but physically interesting case,\cite{YGRS}
of just one family of neutrinos.  Further, let us imagine $m_T=0$ and
$m_S \gg m_D$.  The $2\times 2$ matrix $M$  in this case reads simply
\begin{equation}
M = \left(
\begin{array}{cc}
0 & m_D \\ m_D & m_S 
\end{array} \right)~.
\end{equation}
This matrix
has two eigenvalues, given approximately by $m_S$ and $-m_D^2/m_S$.  That is,
in this case the spectrum splits into a very heavy neutrino of 
(approximate) mass $m_S$ and a very light neutrino of (approximate) mass
$m_D^2/m_S$.\footnote{For fermion fields, the sign of the mass term is
irrelevant since it can be changed by a chiral transformation
$\psi_{\rm R} \to \exp\left[i\frac{\pi}{2}\right]\psi_{\rm R};~\psi_{\rm L}\to
\exp\left[-i\frac{\pi}{2}\right]\psi_{\rm L}$ which leaves the rest of the
Lagrangian invariant.}  This, so called, {\bf see-saw mechanism} is very
suggestive.  It is natural to expect that $m_D$ should be of the order of
the charged lepton mass, corresponding to the neutrino in question:
$m_D\sim m_\ell$.  Then the spectrum of leptons has a natural hierarchy:
\begin{equation}
(m_\nu)_{\rm light} \sim m_\ell\left(\frac{m_\ell}{m_S}\right) \ll m_\ell \ll
(m_\nu)_{\rm heavy} \sim m_S~.
\end{equation}
So, if there is a large mass scale associated with the right-handed neutrinos
(the mass scale $m_S$, which is {\bf not} constrained by the scale of
$SU(2)\times U(1)$ breaking, since it is an $SU(2)\times U(1)$ singlet)
one readily understands why neutrino masses could be so much lighter than
the corresponding charged lepton masses.

The matrix $M$ in the simple $2\times 2$ example above is diagonalized
(approximately) by the orthogonal matrix
\begin{equation}
U = \left(
\begin{array}{cc}
1 & m_D/m_S \\ -m_D/m_S & 1 
\end{array} \right)~.
\end{equation}
The two neutrino mass eigenstates are then
\begin{eqnarray}
\eta_{\rm L} &\equiv& \left(
\begin{array}{c}
\eta_1 \\ \eta_2
\end{array} \right)_{\rm L} = \left(
\begin{array}{cc}
1 & -m_D/m_S \\ m_D/m_S & 1
\end{array} \right) \left(
\begin{array}{c}
\nu_{\rm L} \\ \nu_{\rm R}^c
\end{array} \right) \\
\eta_{\rm R} &\equiv& \left(
\begin{array}{c}
\eta_1 \\ \eta_2
\end{array} \right)_{\rm R} = \left(
\begin{array}{cc}
1 & -m_D/m_S \\ m_D/m_S & 1
\end{array} \right) \left(
\begin{array}{c}
\nu_{\rm L}^c \\ \nu_{\rm R} 
\end{array} \right)~. 
\end{eqnarray}
I note that the mass eigenstates $\eta_1$ and $\eta_2$ are Majorana
(self-conjugate) states
\begin{eqnarray}
\eta_1 &=& \eta_{\rm 1L} + \eta_{\rm 1R} =
(\nu_{\rm L} + \nu_{\rm L}^c) - \frac{m_D}{m_S}
(\nu_{\rm R}^c+\nu_{\rm R}) = \eta_1^c \\
\eta_2 &=& \eta_{\rm 2L} + \eta_{\rm 2R} = 
(\nu_{\rm R}^c + \nu_{\rm R}) + \frac{m_D}{m_S}
(\nu_{\rm L} + \nu_{\rm L}^c) = \eta_{\rm L}^c~.
\end{eqnarray}
The $\nu_{\rm L}$ state which enters in the weak interactions, for all
practical purposes is, essentially $\eta_{\rm 1L}$.  That is, it is the state
associated with the light neutrino eigenstate $(m_1 \simeq m_D^2/m_S)$:
\begin{equation}
\nu_{\rm L} = \eta_{\rm 1L} + \frac{m_D}{m_S} \eta_{2L}~.
\end{equation}
The right-handed neutrino $\nu_{\rm R}$, on the other hand, is essentially
the heavy neutrinos eigenstate $\eta_{2R}~ (m_2\simeq  m_S)$:
\begin{equation}
\nu_{\rm R} = \eta_{\rm 2R} - \frac{m_D}{m_S} \eta_{\rm 1R}~.
\end{equation}

This simple example can be easily generalized to the $3\times 3$ case
of interest.
Again, if the matrix $m_T$ is negligible (i.e. if its eigenvalues are
negligibly small), then the neutrino mass matrix $M$ takes the approximate form
\begin{equation}
M = \left(
\begin{array}{cc}
0 & m_D^T \\ m_D & m_S
\end{array} \right)~.
\end{equation}
Provided the eigenvalues of $m_S$ are large compared to those of $m_D$, then
again the spectrum separates into a light and heavy neutrino sector.  The
light neutrinos have a $3\times 3$ mass matrix
\begin{equation}
(M_\nu)_{\rm light} = m_D^T m_S^{-1} m_D~,
\end{equation}
while the heavy neutrino mass eigenstates are the eigenstates of the
$3\times 3$ matrix
\begin{equation}
(M_\nu)_{\rm heavy} = m_S~.
\end{equation}

The see-saw mechanism, in my view, is the only natural way to understand
eV neutrino masses.  Let me expand a bit on this point.  Simce $m_S$ is an
$SU(2)\times U(1)$ invariant parameter, there are no constraints on it.
On the other hand, as we discussed earlier, both $m_D$ and $m_T$ can only
originate {\bf after} $SU(2)\times U(1)$ breaking.  The Yukawa interaction,
of $\nu_{\rm R}$ with a left-handed doublet ${\rm L} = \left(\begin{array}{c}
\nu_{\ell} \\ \ell
\end{array} \right)_{\rm L}$ via a Higgs doublet $\Phi= \left(\begin{array}{c}
\phi^0\\ \phi^-
\end{array} \right)$
\begin{equation}
{\cal{L}}_{\rm Yukawa} = -\Gamma\bar\nu_{\rm R} \Phi {\rm L} + {\rm h.c.}~,
\end{equation}
leads to a Dirac mass
\begin{equation}
m_D = \Gamma\langle\phi^0\rangle~.
\end{equation}
Since $\langle\phi^0\rangle$ is fixed by the scale of the 
$SU(2)\times U(1)$ breakdown:
\begin{equation}
\langle\phi^0\rangle = \frac{1}{(\sqrt{2}~G_F)^{1/2}} \sim 180~{\rm GeV}~,
\end{equation}
to get $m_D$ to have a value in the eV range requires that
$\Gamma \sim 10^{-11}$!

The situation is not much less artificial in the case of $m_T$.  In this
case, to get a non-zero value for $m_T$ it is necessary to introduce a Higgs
triplet field $\vec\Delta$.  This field can couple to $L\otimes L$ so that if, indeed, $\vec\Delta$ gets a VEV one 
can generate a triplet mass $m_T$.  In detail, the
triplet coupling involving $\vec\Delta$ has the form
\begin{equation}
{\cal{L}}_{\rm triplet} = -\frac{1}{2}\{\Gamma_T L^T\tilde C\vec\tau\cdot
\vec \Delta L\} + {\rm h.c.}~,
\end{equation}
where $\Gamma_T$ is an unknown coupling constant.  When the neutral
component of $\vec\Delta$, $\Delta^o$, gets a vacuum expectation, then
${\cal{L}}_{\rm triplet}$ generates a mass term for $\nu_{\rm L}$:
\begin{equation}
{\cal{L}}^\nu_{\rm mass} = -\frac{1}{2}
\Gamma_T\langle\Delta^o\rangle \{\nu_{\rm L}^T\tilde C\nu_{\rm L}\} +
{\rm h.c.}
\end{equation}
and $m_T = \Gamma_T\langle\Delta^o\rangle$.  The only real constraint on
$\langle\Delta^o\rangle$ comes from precision measurements of the $\rho$ parameter,
typifying the NC to CC ratio.  Experimentally\cite{LEPEW} one finds
\begin{equation}
\rho_{\rm exp} = 1.00412 \pm 0.00124~.
\end{equation}
The presence of the triplet Higgs interaction modifies the $\rho$ parameter
from unity at the tree level and one has:\cite{GR}
\begin{equation}
\rho = 1-2\left(\frac{\langle\Delta^o\rangle}{\langle\phi^0\rangle}\right)^2 +
\mbox{rad. corr.}
\end{equation}
Using the error on $\rho$ in Eq. (62) 
as an estimate of the size of $\langle\Delta^o\rangle$
implies that $\langle\Delta^o\rangle \leq 4$ GeV.  So, also in this case, if $\langle\Delta^o\rangle$ is near this limit to get neutrino masses in the
eV range one needs a
Yukawa strength of order $\Gamma_T \sim 10^{-9}$. If $\langle\Delta^o\rangle
<< \langle \phi^0 \rangle$ then $\Gamma_T$ can be larger, but one is left to explain the reason for the doublet-triplet VEV hierarchy.

Elementary Higgs triplets do not emerge very naturally in models. \footnote{ An exception is provided by left-right symmetric models where triplets have often been considered to give the requisite symmetry breaking. \cite{MMP}}
However, one can always get an {\bf effective} triplet out of two
Higgs doublets:
\begin{equation}
\vec\Delta\sim\Phi^T C \vec\tau\Phi
\end{equation}
where $C$ is an appropriate charge conjugation matrix.  Effective
L-violating interactions involving pairs of 
doublet Higgs fields arise quite naturally in Grand Unified
Theories\cite{Weinberg} as dimension 5 terms:
\begin{equation}
{\cal{L}}_{\rm eff}^{d=5} = \frac{g}{2\Lambda}
(L^T\tilde C\vec\tau L)\cdot(\Phi^TC\vec\tau\Phi) + {\rm h.c.}
\end{equation}
In the above, $\Lambda$ is a scale associated with the GUT breakdown scale and $g$ is
a coupling constant.  Clearly the above interaction gives
\begin{equation}
m_T =  \frac{g\langle\phi^0\rangle^2}{\Lambda}~.
\end{equation}
Since $\langle\phi^0\rangle\sim 10^2$ GeV, with $g\sim O(1)$, one gets
$m_T \sim 10^{-2}$ eV for scales $\Lambda \sim 10^{15}$ GeV, which are typical of GUTs.  Note that
the above formula for $m_T$ is quite similar in spirit to the see-saw expression for
light neutrinos
\begin{equation}
(m_\nu)_{\rm light}^{\rm see-saw} \sim \frac{m_D^2}{m_S} \sim
\frac{\langle\phi^0\rangle^2}{m_S}~,
\end{equation}
since $m_D \sim\langle\phi^0\rangle$.  In either case, new physics at a large
scale (either a large $\nu_{\rm R}$ mass scale $m_S$ or the GUT scale
$\Lambda$) produces a light neutrino.  It is clearly more appealing physically
to have light neutrinos be the result of new physics at high scales, rather
than simply as a result of some Yukawa coupling being unnaturally small.

\section{Neutrino Oscillations in Vacuum}

If neutrinos have mass then, in general, the neutrinos produced by the
weak interactions (weak interaction eigenstates) are not states of
definite mass (mass eigenstates).  In the basis where the charged
lepton mass matrix is diagonal [c.f. Eq. (25)], it follows from 
Eq. (23) that the neutrino weak interaction eigenstates are fixed by
the corresponding lepton produced in the associated  weak process.  That is, the
piece of the weak current $J_{\mu -}^{\rm lept}$ involving the charged
lepton $\ell_i = \{e,\mu,\tau\}$ will {\bf always} involve the corresponding neutrino
weak interaction eigenstate $\nu_{\ell_i} = \{\nu_e,\nu_\mu,\nu_\tau\}$.
These, left-handed, neutrino weak interaction eigenstates are superpositions of neutrino mass eigenstates $\nu_i$:
\begin{equation}
\nu_{\ell_j} = \sum_i U_{\ell_ji}\nu_i~.
\end{equation}
The matrix $U_{\ell_ji}$, in general, is a $3\times 6$ matrix.  However, if the see-saw mechanism is operative,
one expects that the contributions of {\bf superheavy} 
neutrinos in Eq. (68) should
be negligible.  Then, to a very good approximation, the matrix
$U_{\ell_ji}$ is a $3\times 3$ unitary matrix
\begin{equation}
U_{\ell_ji} U^*_{\ell_ki} = \delta_{\ell_j\ell_k}~.
\end{equation}

For the moment I will restrict myself to the case when Eq. (69) holds. Furthermore, I will discuss the phenomenology of neutrino oscillations in
the simple case of just 2 flavors of neutrinos, since this is how most of the data is usually presented.  However, the formalism  which  we will develop can 
be generalized straightforwardly to three families of light neutrinos. \cite{reviews} For
definitiveness, let us consider then just $\nu_e$ and $\nu_\mu$ weak
interaction eigenstates.  In this case, Eq. (68) reads, using a
convenient quantum mechanical notation,
\begin{eqnarray}
|\nu_e\rangle &=& \cos\theta|\nu_1\rangle + \sin\theta|\nu_2\rangle 
\nonumber \\
|\nu_\mu\rangle &=& -\sin\theta|\nu_1\rangle + \cos\theta
|\nu_2\rangle~. 
\end{eqnarray}
The mass eigenstates $|\nu_i\rangle$ have a time evolution which just
follows from the Schr\"odinger equation:
\begin{equation}
|\nu_i(t)\rangle = e^{-iE_it}|\nu_i(0)\rangle~; ~~~
E_i = \sqrt{\vec p^2 + m_i^2}~.
\end{equation}

Because $m_1\not= m_2$, it is easy to see that the weak interaction
eigenstate $\nu_e$ produced at $t=0$ evolves in time into a
superposition of $\nu_e$ and $\nu_\mu$ states.  Taking by definition
$|\nu_i\rangle \equiv |\nu_i(0)\rangle$, it follows that
\begin{eqnarray}
|\nu_e(t)\rangle &=& \cos\theta e^{-iE_1t}|\nu_1\rangle +
\sin\theta e^{-iE_2t}|\nu_2\rangle \nonumber\\
&=&\left[\cos^2\theta e^{-iE_1t} + \sin^2\theta e^{-iE_2t}\right]|\nu_e\rangle +
\left[\cos\theta\sin\theta(e^{-iE_2t}-e^{-iE_1t})\right]
|\nu_\mu\rangle \nonumber \\
&\equiv& A_{ee}(t)|\nu_e\rangle + A_{e\mu}(t)|\nu_\mu\rangle~.
\end{eqnarray}

Using the above, one can compute immediately the probabilities that
at time $t$ the state $\nu_e(t)$ is either a $\nu_e$ or a $\nu_\mu$ weak
interaction eigenstate:
\begin{eqnarray}
P(\nu_e\to\nu_e;t) &=& |A_{ee}(t)|^2 = 1-\frac{1}{2}\sin^22\theta
[1-\cos(E_2-E_1)t] \\
P(\nu_e\to\nu_\mu;t) &=& |A_{e\mu}(t)|^2 = \frac{1}{2}\sin^22\theta
[1-\cos(E_2-E_1)t]~.
\end{eqnarray}
Since the masses of neutrinos are small compared to the momentum, one
can write
\begin{equation}
E_i \simeq |p| + \frac{m_i^2}{2|p|}~; ~~~~ t\simeq L~,
\end{equation}
where $L$ is the distance travelled by the neutrinos in a time $t$.
Using the above, one can write, for instance,
\begin{equation}
P(\nu_e\to\nu_\mu;L) = \frac{1}{2}sin^22\theta
\left[1-\cos\frac{\Delta m^2}{2|p|} L\right] =
\sin^22\theta\sin^2\frac{\Delta m^2L}{4|p|}~,
\end{equation}
where $\Delta m^2 = m_2^2-m_1^2$.  Numerically, it turns out that
\begin{equation}
\frac{\Delta m^2L}{4|p|} \simeq 1.27 
\frac{\Delta m^2({\rm eV}^2) L(m)}{|p|({\rm MeV})}~.
\end{equation}

Recapitulating, for the case of 2 neutrino species, one gets the
following formula quantifying the probability that a weak interaction
eigenstate neutrino $(\nu_e)$ has oscillated to other weak
interaction eigenstate neutrino $(\nu_\mu)$ after traversing a
distance $L$:
\begin{equation}
P(\nu_e\to\nu_\mu;L) = \sin^22\theta\sin^2
\left[\frac{1.27 \Delta m^2({\rm eV}^2) L(m)}{|p| ({\rm MeV})}\right]~.
\end{equation}
The probability that no such oscillation took place, of course, is
just
\begin{equation}
P(\nu_e\to\nu_e;L) = 1 - P(\nu_e\to\nu_\mu;L)~.
\end{equation}
These probabilities depend on two factors: (i) a mixing angle factor
$\sin^22\theta$ and (ii) a kinematical factor which depends on the
distance travelled, on the momentum of the neutrinos, as well as on
the difference in the squared mass of the two neutrinos.  Obviously, for oscillations to
be important the mixing factor $\sin^22\theta$
should be of $O(1)$.  However,
large mixing is not enough. It is also important that the kinematical
factor $\Delta m^2({\rm eV}^2) L(m)/|p| ({\rm MeV}) 
\stackrel{>}{_{\scriptstyle \sim}} O(1)$, so that the second oscillatory
factor in Eq. (78) can be significant.

It is useful to develop the formalism a bit more for future use.  The
probability amplitudes $A_{ee}(t)$ and $A_{e\mu}(t)$ of Eq. (72) can
be recognized as matrix elements of a $2\times 2$ matrix $e^{-iHt}$
defined by
\begin{equation}
e^{-iHt} = U~e^{-iH_{\rm diag}t}~U^\dagger~,
\end{equation}
where
\begin{equation}
U = \left(
\begin{array}{cc}
\cos\theta & \sin\theta \\ -\sin\theta & \cos\theta
\end{array} \right)
\end{equation}
is the mixing matrix of the weak interaction eigenstates and
\begin{equation}
H_{\rm diag} = \left(
\begin{array}{cc}
E_1 & 0 \\ 0 & E_2 
\end{array} \right)~.
\end{equation}
One has
\begin{equation}
A_{ee}(t) = [e^{-iHt}]_{11}~; ~~~~
A_{e\mu}(t) = [e^{-iHt}]_{12}~.
\end{equation}
Using the fact that $E_i = |p| + m_i^2/2|p|$, it proves convenient to
separate $H_{\rm diag}$ into two different pieces:
\begin{equation}
H_{\rm diag} = \left(|p| + \frac{m_1^2 + m_2^2}{4|p|}\right) \left(
\begin{array}{cc}
1 & 0 \\ 0 & 1
\end{array} \right) +
\frac{\Delta m^2}{4|p|} \left[
\begin{array}{cc}
-1 & 0 \\ 0 & 1
\end{array} \right]~.
\end{equation}
The first piece above, because it is proportional to the unit matrix,
gives an overall phase factor which is irrelevant for
calculating the neutrino oscillation probabilities.  Hence, effectively,
one can replace
\begin{equation}
H_{\rm diag} \to H_o = -\frac{\Delta m^2}{4|p|} \sigma_3~.
\end{equation}
In view of (85), it is convenient to define the $2\times 2$
Hamiltonian matrix $H_{\rm vac}$ by
\begin{eqnarray}
H_{\rm vac} &=& U~H_o~U^\dagger = \frac{\Delta m^2}{4|p|} \left(
\begin{array}{cc}
-\cos2\theta & \sin2\theta \\
\sin2\theta & \cos2\theta
\end{array} \right) \nonumber \\
&=& \frac{\Delta m^2}{4|p|} \{\sin2\theta\sigma_1 -
\cos2\theta\sigma_3\}~.
\end{eqnarray}
Then, effectively,
\begin{equation}
A_{ee}(t) = [e^{-iH_{\rm vac}t}]_{11}~; ~~~~
A_{e\mu}(t) = [e^{-iH_{\rm vac}t}]_{12}~.
\end{equation}

Just as the above coefficients describe the time evolution of a state
that started at $t=0$ as a $\nu_e$ weak interaction eigenstate
[cf. Eq. (72)], one can define coefficients
\begin{equation}
A_{\mu e}(t) = [e^{-iH_{\rm vac}t}]_{21}~; ~~~~
A_{\mu\mu}(t) = [e^{-iH_{\rm vac}t}]_{22}
\end{equation}
which will detail the time evolution of a state which started out at
$t=0$ as a $\nu_\mu$ weak interaction eigenstate:
\begin{equation}
|\nu_\mu(t)\rangle = A_{\mu e}(t)|\nu_e\rangle +
A_{\mu\mu}(t)|\nu_\mu\rangle~.
\end{equation}
It is easy to deduce from these considerations that the $2\times 2$
Hamiltonian $H_{\rm vac}$ is just the Hamiltonian which enters in
the Schr\"odinger equation for $|\nu_e(t)\rangle$ and
$|\nu_\mu(t)\rangle$:
\begin{equation}
i\frac{\partial}{\partial t} \left[
\begin{array}{c}
|\nu_e(t)\rangle \\ |\nu_\mu(t)\rangle
\end{array} \right] = H_{\rm vac} \left[
\begin{array}{c}
|\nu_e(t)\rangle \\ |\nu_\mu(t)\rangle
\end{array} \right]~.
\end{equation}

\section{Neutrino Oscillations in Matter}

When neutrinos propagate in matter, a subtle but important effect
takes place which alters the ways in which neutrinos oscillate into
one another.  The origin of this effect, which is known as the MSW
effect for the initials of the physicists who first discussed it,\cite{MSW} is connected to the fact that the electron neutrinos
can interact in matter also through charged current interactions.
While all neutrino species have the same 
interactions in matter due to the neutral currents, the $\nu_e$ weak
interaction eigenstates, because of their charged current interactions, as they propagate in matter experience a
slightly different index of refraction than the $\nu_\mu$ weak
interaction eigenstates (and the $\nu_\tau$ weak interaction eigenstates).  This different index
of refraction for $\nu_e$ alters the time evolution of the system from
what it was in vacuum.

Let us again consider the two-neutrino case.  The relative index of
refraction between $\nu_e$ and $\nu_\mu$ is the result of the difference
between the forward scattering amplitudes for $\nu_e$ and $\nu_\mu$, caused by the charged current interactions of the $\nu_e$.  In detail,
one has\cite{MSW}
\begin{equation}
1-n_{\rm rel} = -\frac{2\pi N_e}{|p|^2} \left[\left.A(0)\right|_{\nu_ee}
-\left.A(0)\right|_{\nu_\mu e}\right]
\end{equation}
where $N_e$ is the electron density and $A(0)$ is the forward
scattering amplitude.  The contribution of the neutral current
interactions cancels in Eq. (91), while the charged current
contribution to $\left.A(0)\right|^{\rm CC}_{\nu_ee}$ gives\cite{MSW}
\begin{equation}
1-n_{\rm rel} = \frac{\sqrt{2} G_F~N_e}{|p|}~.
\end{equation}

One can use Eq. (92) and the formalism we developed at the end of the
last section to study the evolution of neutrinos in matter.  The
relevant Hamiltonian now is
\begin{equation}
H = H_{\rm vac} + |p|(1-n_{\rm rel}) \left[
\begin{array}{cc}
1 & 0 \\ 0 & 0
\end{array} \right]~.
\end{equation}
Again, because relative phases are irrelevant, we can subtract from
the above a term proportional to the identity.  This yields the
following effective Hamiltonian describing the propagation of neutrinos
in matter:
\begin{eqnarray}
H_{\rm matter} &=& H_{\rm vac} + |p| \frac{(1-n_{\rm rel})}{2} \left[
\begin{array}{cc}
1 & 0 \\ 0 & -1 
\end{array} \right] \nonumber \\
&=& \frac{\Delta m^2}{4|p|} \sin 2\theta \sigma_1 - \left(
\frac{\Delta m^2}{4|p|} \cos2\theta - \frac{G_F~N_e}{\sqrt{2}}\right)
\sigma_3~.
\end{eqnarray}

Because of the term proportional to the electron density in Eq. (94),
in matter it is no longer true that the eigenstates of $H_{\rm matter}$
are $\nu_1$ and $\nu_2$.  Calling these matter eigenstates
$\nu_1^M$ and $\nu_2^M$, one has that
\begin{equation}
\left[
\begin{array}{c}
|\nu_e\rangle \\ |\nu_\mu\rangle
\end{array} \right] = \left(
\begin{array}{cc}
\cos\theta_M & \sin\theta_M \\ -\sin\theta_M & \cos\theta_M 
\end{array} \right) \left[
\begin{array}{c}
|\nu_1^M\rangle \\ |\nu_2^M\rangle
\end{array} \right] \equiv U_M \left[
\begin{array}{c}
|\nu_1^M\rangle \\ |\nu_2^M\rangle
\end{array} \right]
\end{equation}
and
\begin{equation}
H_{\rm matter} = U_M H^{\rm diag}_{\rm matter} U_M^\dagger~.
\end{equation}
It is easy to check that $H^{\rm diag}_{\rm matter}$ is given by
\begin{equation}
H^{\rm diag}_{\rm matter} = -\sigma_3\left[\left(\frac{\Delta m^2}
{4|p|} \cos2\theta - \frac{G_F~N_e}{\sqrt{2}}\right)^2 + \left(
\frac{\Delta m^2}{4|p|} \sin2\theta\right)^2\right]^{1/2}~,
\end{equation}
with the mixing angle in matter $\theta_M$ determined by the
equation
\begin{equation}
\sin2\theta_M = 
\frac{\frac{\Delta m^2}{4|p|}\sin2\theta}
{\left[\left(\frac{\Delta m^2}{4|p|}\cos2\theta -
\frac{G_F~N_e}{\sqrt{2}}\right)^2 + \left(
\frac{\Delta m^2}{4|p|}\sin2\theta\right)^2\right]^{1/2}}~.
\end{equation}

The presence of the term proportional to the electron density gives rise
to interesting resonance phenomena.\cite{Petcov}  There is a {\bf critical
density} $N_e^{\rm crit}$, given by
\begin{equation}
N_e^{\rm crit} = \frac{\Delta m^2\cos2\theta}{2\sqrt{2}|p|G_F}~,
\end{equation}
for which the matter mixing angle $\theta_M$ becomes {\bf maximal}
$(\sin2\theta_M \to 1)$, irrespective of what the vacuum mixing angle
$\theta$ is.  If one is in such a medium, then $H^{\rm diag}_{\rm matter}$ reduces to
\begin{equation}
\left.H^{\rm diag}_{\rm matter}\right|_{N_e=N_e^{\rm crit}} = -
\frac{\Delta m^2}{4|p|} \sin2\theta\sigma_3~.
\end{equation}
The probability that a $\nu_e$ transmutes into a $\nu_\mu$ after
traversing a distance $L$ in this medium is given by Eq. (78), with
two differences.  First, since we are in a medium $\sin2\theta\to
\sin2\theta_M$.  However, because the density is assumed to be the critical density,
$\sin2\theta_M\to 1$.  Second, since $H^{\rm diag}_{\rm matter}$ in
Eq. (100) differs 
from $H_o$ by the replacement of $\Delta m^2\to\Delta m^2\sin2\theta$, such a replacement also will enter in the kinematical
factor in the probability formula.  Hence, it follows that
\begin{equation}
P_{\rm matter} \left.(\nu_e\to\nu_\mu;L)\right|_{N_e=N_e^{\rm crit}} =
\sin^2\left(\frac{\Delta m^2}{4|p|}\sin2\theta L\right)~.
\end{equation}
This formula shows that one can get {\bf full conversion} of a $\nu_e$
weak interaction eigenstate into a $\nu_\mu$ weak interaction
eigenstate, provided that the length $L$ and momentum $|p|$ satisfy
the relation
\begin{equation}
\frac{\Delta m^2}{4|p|} \sin2\theta L = \frac{n\pi}{2}~; ~~~~
n = 1,2,\ldots~.
\end{equation}

There is a second interesting limit to consider. \cite{Petcov}  This is when the
electron density $N_e$ is so large that it overwhelms the other terms
in $H^{\rm diag}_{\rm matter}$.  If $G_F~N_e \gg \Delta m^2/2\sqrt{2}|p|$, then one has, approximately,
\begin{equation}
H^{\rm diag}_{\rm matter} = -\sigma_3 \frac{G_F~N_e}{\sqrt{2}}~.
\end{equation}
In this limit, it is easy to check that $\sin 2\theta_M\to 0$;
$\cos2\theta_M\to -1$, so that $\theta_M\to \frac{\pi}{2}$.  In this
case, there are no oscillations in matter because $\sin2\theta_M$
vanishes
\begin{equation}
P_{\rm matter}\left.(\nu_e\to\nu_\mu;L)\right|_{N_e \gg \frac{\Delta m^2}{2\sqrt{2}|p|G_F}} \to 0~.
\end{equation}
This actually is immediate also since, in this limit, $H_{\rm matter}$
itself is diagonal
\begin{equation}
H_{\rm matter} = U_M H^{\rm diag}_{\rm matter} U^\dagger_M =\sigma_3
\frac{G_F~N_e}{\sqrt{2}}~.
\end{equation}
Hence the Schr\"odinger equation (90), with $H_{\rm vac}\to H_{\rm matter}$, is diagonal and there can be no transitions.  For
future use I note that in this limit, since $\theta_M = \pi/2$, the
$\nu_e$ weak interaction eigenstate in matter coincides with the
state $\nu_2^M$:
\begin{equation}
|\nu_e\rangle = \cos\theta_M|\nu_1^M\rangle + 
\sin\theta_M|\nu_2^M\rangle \stackrel{\rightarrow}{_{\theta_M=\pi/2}}
|\nu_2^M\rangle~.
\end{equation}

\section{Evidence for Neutrino Masses}

Most experiments searching for direct evidence for neutrino masses have, up
to now, only set limits on these masses and the associated mixing angles.
However, there are now both strong hints, and some real evidence, 
that neutrino
masses really exist coming from neutrino oscillation experiments.

Most oscillation data is presented as an allowed region, or limits at some
confidence level, in a $\Delta m^2-\sin^22\theta$ plot.  That is, experimentalists find it convenient to quantify their results using the 
formalism discussed in the last section, involving oscillations among two
neutrino species $\nu_\alpha$ and $\nu_\beta$  
The oscillation probability formulas for
$P(\nu_\alpha\to\nu_\beta;L)$ [cf. Eq. (78)] involves both the $\alpha-\beta$
mixing angle $\theta_{\alpha\beta}$, as well as a kinematical factor depending 
on the mass squared difference $\Delta m^2$ between the mass eigenstates in the two neutrino system.
Because neutrino beams have a rather large energy spread, for 
$\Delta m^2 \gg |p|/L$ the kinematical oscillating factor in Eq. (78) averages
to 1/2.  This implies that, in general, the sensitivity to a signal for
neutrino oscillations goes down to \break $\sin^22(\theta_{\alpha\beta})_{\rm min} \simeq 2 
P(\nu_\alpha\to \nu_\beta;L)$.

\subsection{Direct Mass Measurements}

The classical way to try to infer a non-vanishing value for neutrino masses
is by measuring $\beta$-decay spectra near their endpoint.  The presence
of neutrino masses alters the dependence of the measured intensity 
$I(T_e)$ on the electron kinetic energy $T_e$ as it approaches the
maximum energy release $Q$.  One has \cite{Kurie}
\begin{equation}
I(T_e) = (Q-T_e) \sum_i |U_{ei}|^2
[(Q-T_e)^2-m^2_{\nu_i}]^{1/2}~.
\end{equation}
If $m_{\nu_i} = 0$ then the intensity spectrum is quadratically dependent
on the energy release $(Q-T_e)$.  If neutrinos have mass, one has a spectrum
distorsion and $\sqrt{I}$ is no longer linear in $(Q-T_e)$, but vanishes at some
value of $T_e$ less than the maximum energy released $Q$.  These distorsions
are best detected in $\beta$-decay spectra which have low $Q$ values; an
ideal candidate being Tritium where $Q = 18.6$ KeV.

Tritium $\beta$-decay experiments are sensitive to neutrino masses in the
``few eV" range.  Remarkably, most of the high precision experiments performed with tritium
actually see an {\bf excess} of events near the end-point, setting 
poorer limits than their theoretical sensitivity.\cite{Zuber}  However,
very recently, the Troitsk experiment \cite{Troitsk} has been able to determine
a very stringent result for the largest eigenvalue ``$m_{\nu_e}$" principally
contributing to Tritium $\beta$-decay:
\begin{equation}
``m^2_{\nu_e}" = (1.5\pm 5.9 \pm 3.6)~{\rm eV}^2~; ~~~
``m_{\nu_e}" < 3.9~{\rm eV}~~(90\%~{\rm C.L.)}
\end{equation}
Table 1 gives a compilation of the existing $\beta$-decay results in
Tritium and the corresponding limit for ``$m_{\nu_e}$".

\begin{table}
\caption{Neutrino Mass Limits from  {${}^3$He $\beta$}-decay, from
Ref. [14].}  
\label{Table1}
\begin{tabular}{ccc} 
Experiment & ``$m^2_{\nu_e}$"(eV$^2$) & ``$m_{\nu_e}$"(eV)\\ 
~&~&~\\
Tokyo  & $-65\pm 85\pm 65$ & $<13.1$ \\
Los Alamos & $-147\pm 68\pm 41$ & $<9.3$ \\
Z\"urich & $-24\pm 48\pm 61$ & $<11.7$ \\
Livermore & $-130\pm 20\pm 15$ & $<7.0$ \\
Mainz & $-22\pm 17\pm 14$ & $<5.6$ \\
Troitsk & $1.5\pm 5.9\pm 3.6$ & $<3.9$ \\ 
\end{tabular}
\end{table}

Similar, but less accurate, kinematical bounds are also known for the
largest eigenvalues principally contributing to decays involving
$\nu_\mu$ and $\nu_\tau$ weak eigenstates.  Denoting these eigenvalues,
respectively, as ``$m_{\nu_\mu}$" and ``$m_{\nu_\tau}$", one finds the
following results. From studying $\pi^+\to \mu^+\nu_\mu$ decay at
PSI \cite{PSI} one has
\begin{equation} 
``m_{\nu_\mu}^2" = (-0.016\pm 0.028)~ {\rm MeV}^2;~~ 
``m_{\nu_\mu}" < 170~ {\rm KeV}~ (90\% C.L.)~.
\end{equation} 
From studying the decay
$\tau\to\nu_\tau + 5\pi$ at LEP\cite{ALEPHT} one has 
\begin{equation}
``m_{\nu_\tau}" < 18.2~{\rm MeV}~ (95\% ~{\rm C.L.})~.
\end{equation}

A different, and in some ways more interesting, limit on the neutrinos
associated with the $\nu_e$ weak interaction eigenstate comes from neutrinoless
double $\beta$-decay.  This process, if it exists, violates lepton number.
Thus it is only possible if neutrinos have a Majorana mass. 
Ordinary double $\beta$-decay $Z\to (Z+2) + 2e^- + 2\bar\nu_e$ 
conserves lepton number.  However, in a 
double $\beta$-decay processes where no
neutrinos are emitted $Z\to (Z+2) + 2e^-$, lepton number is violated.  
As shown
schematically in Fig. 1, these processes can only occur if there is a
neutrino-antineutrino transition engendered by the presence of a Majorana
mass term.

\begin{figure}
\centerline{\epsfig{file=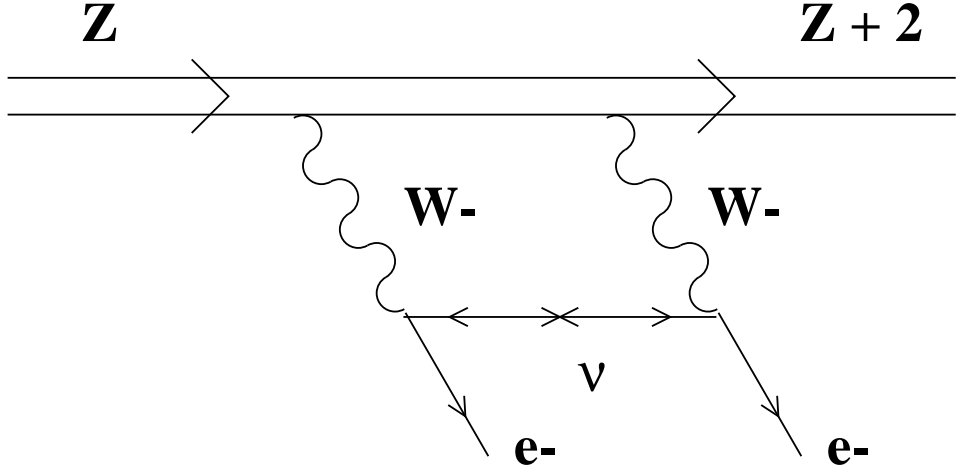,width=5in}}
\caption[]{Schematic diagram which gives rise to neutrinoless double  {$\beta$}-decay if neutrinos have a Majorana mass.}
\end{figure}

There is now a variety of measurements of ordinary double $\beta$-decay,\cite{2beta} but up to now there are only limits on neutrinoless
double $\beta$-decay. \cite{0beta}  The half-life for these later processes is a measure
of the neutrino Majorana mass associated with these decays:
\begin{equation}
\left[T_{\frac{1}{2}}^{2\beta o\nu}\right]^{-1} \sim
\langle m_{\nu_e}\rangle_{ee}^2~.
\end{equation}
Here the mass $\langle m_{\nu_e}\rangle_{ee}$ is given by
\begin{equation}
\langle m_{\nu_e}\rangle_{ee} = \sum_i U_{ei}^2 m_{\nu_i},
\end{equation}
with $m_{\nu_i}$ being the  neutrino mass eigenstates.  Note that $\langle m_{\nu_e}\rangle_{ee}$ vanishes if the neutrinos are Dirac particles.
That is, if neutrinos have only lepton number conserving Dirac masses.

This point can be appreciated readily by considering,
for simplicity, the case of 
one neutrino species.
In this case, if these neutrinos only have a Dirac mass
$m_D$, the corresponding $2\times 2$  neutrino mass matrix $M$ has the form
\begin{equation}
M = \left(
\begin{array}{cc}
0 & m_D \\ m_D & 0
\end{array} \right)~.
\end{equation}
This matrix is diagonalized to
\begin{equation}
M_{\rm diag} = \left(
\begin{array}{cc}
-m_D & 0 \\ 0 & m_D
\end{array} \right)
\end{equation}
by the orthogonal matrix
\begin{equation}
U = \left(
\begin{array}{cc}
\frac{1}{\sqrt{2}} & \frac{1}{\sqrt{2}} \\
-\frac{1}{\sqrt{2}} & \frac{1}{\sqrt{2}}
\end{array} \right)~.
\end{equation}
Using Eq. (115), for this case one easily checks that $\langle m_{\nu_e}\rangle_{ee}$ vanishes. One has
\begin{equation}
\langle m_{\nu_e}\rangle_{ee} = \left(\frac{1}{\sqrt{2}}\right)^2
(-m_D) + \left(\frac{1}{\sqrt{2}}\right)^2 (m_D) = 0~.
\end{equation}

Table 2 reproduces a  recent compilation of experimental
results on 
neutrinoless double $\beta$-decay. \cite{Zuber}  The most sensitive
of these experiments involves the double $\beta$-decay of ${}^{76}$Ge to
${}^{76}$Se, with a half-life limit of over $10^{25}$ years. \cite{MH}  The resulting
bound on $\langle m_{\nu_e}\rangle_{ee}$ quoted is
\begin{equation}
\langle m_{\nu_e}\rangle_{ee} < 0.2~{\rm eV} ~~~~
(90\%~ {\rm C.L.})~.
\end{equation}
This bound, however, has probably a factor of two uncertainty due to
uncertainties associated with calculating the nuclear matrix elements
involved in the decay.\cite{matrix} 

\begin{table}
{\caption{Bounds on neutrinoless double {$\beta$}-decays half lives and
associated bounds on {$\langle m_{\nu_e}\rangle_{ee}$}, from 
Ref. [14].} }
\label{Table 2}
\begin{tabular}{ccc} 
Decay & $T_{\frac{1}{2}}^{2\beta o\nu}$ (Years) & 
$\langle m_{\nu_e}\rangle_{ee}$ (eV) \\ 
~&~&~\\
${}^{76}{\rm Ge}\to {}^{76}{\rm Se}$ & $>5.7\times 10^{25}$ (90\% C.L.) &
$<0.2$ (90\% C.L.) \\
${}^{128}{\rm Te}\to {}^{128}{\rm Xe}$ & $>7.7\times 10^{24}$ (68\% C.L.) &
$<1.1$ (68\% C.L.) \\
${}^{130}{\rm Te}\to {}^{130}{\rm Xe}$ & $>5.6\times 10^{22}$ (90\% C.L>) &
$<3.0$ (90\% C.L.) \\
${}^{136}{\rm Xe}\to {}^{136}{\rm Ba}$ & $>4.4\times 10^{23}$ (90\% C.L.) &
$<2.3$ (90\% C.L.)\\  
\end{tabular}
\end{table}

\subsection{Cosmological Constraints}

There are some indirect constraints on neutrino masses provided by cosmology.
The most relevant is the constraint which follows from demanding that
the energy density in neutrinos should not overclose the Universe.  
Neutrinos are thermal relics; they decoupled from the Universe's expansion
when their interaction rate $\Gamma$ fell below the Universe's expansion
rate $H$. \cite{KT}  In the usual Robertson-Walker expanding Universe, the rate of
expansion $H\sim T^2/M_{\rm P}$, where $M_{\rm P}\sim 10^{19}$ GeV is the
Planck mass and $T$ is the Universe's temperature.  Since
\begin{equation}
\Gamma = n_\nu \langle\sigma v\rangle\sim G_F^2 T^5~,
\end{equation}
with $G_F$ the Fermi constant, $G_F\sim 10^{-5}~{\rm GeV}^{-2}$, decoupling
occurs at a temperature $T_D$ determined by setting
$\Gamma \simeq H$.  This gives
\begin{equation}
T_D\simeq\left(\frac{1}{G_F^2 M_{\rm P}}\right)^{1/3} \sim 1~{\rm MeV}~.
\end{equation}

Two cases are of interest.  If neutrinos have a mass much less than $T_D$
($m_\nu \ll T_D$) they are {\bf hot relics}.  That is, they are relativistic
at the time of decoupling.  For hot relics, the density of neutrinos is
comparable to that of photons at decoupling: $\left.n_\nu\sim n_\gamma\right|_{T_D}$.
{\bf Cold relics}, on the other hand, are neutrinos whose mass is 
much greater than
$T_D$ ($m_\nu \gg T_D$).  In this case, at the time of decoupling the
neutrino density $n_\nu$, because of the Boltzmann factor, is much below
that of the photons.  Thus for cold relics, $\left.n_\nu \ll n_\gamma\right|_{T_D}$.

In either case, one can compute the neutrino contribution to the energy
density of the Universe. \cite{KT}  This is simplest for the case of hot relics, since
their number density essentially  tracks the photon number density.\footnote{Because of photon reheating at the time of recombination
(and a small statistical difference because neutrinos are fermions and photons are
bosons), the neutrino temperature $T_\nu$ is not quite the same as the photon
temperature $T_\gamma$.  One finds $T_\nu = \left(\frac{4}{11}\right)^{1/3} T_\gamma$.\cite{Weinberg1}}  
Thus the number density of
neutrinos now is fixed by the measured temperature of the microwave background
radiation:
\begin{equation}
n_\nu = \frac{3\zeta(3)}{2\pi^2} T_\nu^3~.
\end{equation}
Hence the contribution of neutrinos to the present energy density of the
Universe is
\begin{equation}
\rho_\nu = n_\nu\sum_i m_{\nu_i}~.
\end{equation}

It has become conventional to normalize all densities in terms of the
Universe's closure density $\rho_c$:
\begin{equation}
\rho_c = \frac{3H_o^2}{8\pi G_N} \simeq 1.9\times 10^{-29} h^2
\frac{\rm g}{\rm cm^3} \simeq 1.1\times 10^4 h^2 \frac{\rm eV}{\rm cm^3}~.
\end{equation}
In the above $H_o$ is the Hubble constant and $h$ is a measure of its
uncertainty.  One finds
\begin{equation}
H_o = 100h \frac{\rm Km}{\hbox{sec Mpsec}}
\end{equation}
with\cite{Hune}
\begin{equation}
h = 0.65\pm 0.1~.
\end{equation}
Defining
\begin{equation}
\Omega_\nu = \frac{\rho_\nu}{\rho_c}~,
\end{equation}
then, for hot relics, one has
\begin{equation}
\Omega_\nu^{\rm Hot} = \frac{\sum_i m_{\nu_i}}{92~{\rm eV}~h^2}~.
\end{equation}
It follows from the above that if the sum of neutrino masses $\sum_i m_{\nu_i} \simeq 30~{\rm eV}$, then
neutrinos would close the Universe.
Because we know that the Universe is not very far from closure density, if
neutrinos are hot relics the sum of their masses cannot be much above
30 eV.  Thus, although direct bounds allows for a presence of a
``$\nu_\mu$" neutrino with mass ``$m_{\nu_\mu}$" less than 170 KeV,
cosmology forbids neutrinos to have masses as large as that. \footnote{These cosmological bounds can be avoided if the massive neutrinos were unstable and had a sufficiently short lifetime. \cite{unstable}}

When neutrino
masses are above $T_D \sim {\rm MeV}$, then the simple formula given in Eq. (126) no longer
applies.  Nevertheless, it is still possible to compute $\Omega_\nu$ taking
into account now of the appropriate Boltzmann factor.
Figure 2, adapted from \cite{KT}, plots $\Omega_\nu h^2$ as a function of the neutrino (sum) mass.
This quantity rises linearly with mass up to $m_\nu \sim 1~{\rm MeV}$ and then
drops rather rapidly.  Cosmology allows neutrino masses for which
$\Omega_\nu \leq 1$.  So, as mentioned above, ``$m_{\nu_\mu}$" and
``$m_{\nu_\tau}$" must really be well below their kinematical bounds.  On the
other hand, we note that our bounds for ``$m_{\nu_e}$" and 
$\langle m_{\nu_e}\rangle_{ee}$ lie in a cosmologically allowed
region.
In principle, cosmology also allows neutrinos to exist with masses greater
than a few GeV, since these cold relics give $\Omega_\nu \leq 1$.  However,
as we discussed earlier, neutrino counting at LEP excludes additional
neutrinos besides $\nu_e,\nu_\mu$ and $\nu_\tau$, with mass $m_\nu < M_Z/2$.
This exclusion region is also indicated in Fig. 2.

\begin{figure}
\epsfig{file=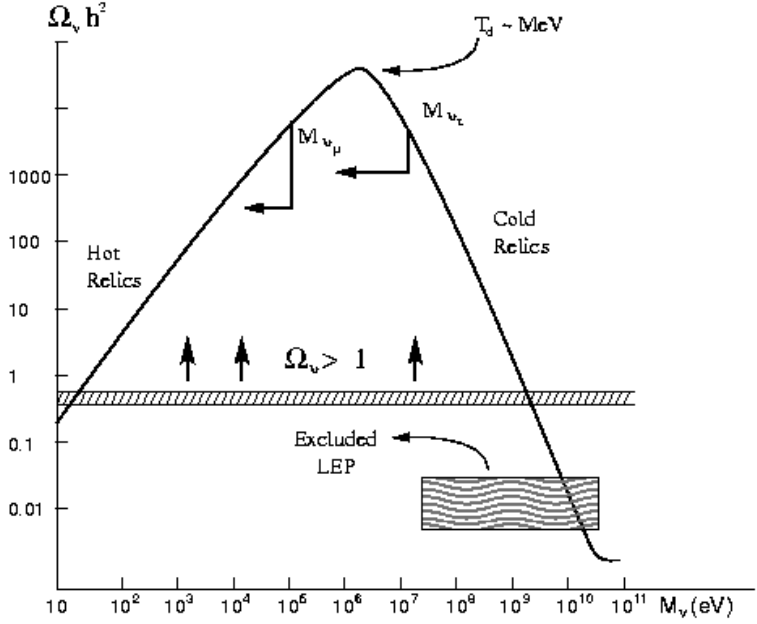,width=5in}
\caption[]{Plot of {$\Omega_\nu h^2$} as a function of the neutrino (sum) mass.}
\end{figure}

\subsection{Accelerator limits and hints for neutrino masses}

Two experiments at CERN, using $\nu_\mu$ beams of average neutrino energies
$\langle E_{\nu_\mu}\rangle\sim 15~{\rm GeV}$ and typical decay lengths
$L \sim 1~{\rm Km}$, put strong limits on $\nu_\mu\to\nu_\tau$ neutrino
oscillations for $\Delta m^2\geq 10~{\rm eV}^2$.  In view of the discussion
in the last subsection, this is a very interesting mass range to explore, since neutrinos with masses in this range could be of cosmological interest.

These two experiments use quite different techniques to detect $\nu_\tau$'s. \break
CHORUS\cite{chorus} uses an emulsion target to try to detect the $\tau$ track
produced in $\nu_\tau$ charged current interactions.  NOMAD, \cite{nomad}
on the other hand, uses drift chambers and kinematical techniques to detect
a $\nu_\tau$ signal.  The result of both experiments are shown in Fig. 3,
along with limits obtained by some earlier experiments.  
The CHORUS and NOMAD results
for $\Delta m^2 \geq 10~{\rm eV}^2$ exclude oscillations with mixing angles
$\sin^22\theta_{\mu\tau} \geq 10^{-3}$ at the 90\% C.L., improving previous
limits by about a factor of 5.

Because NOMAD has a very good electron identification, this experiment
is also able to set a strong limit for $\nu_\mu\to\nu_e$ oscillations.
For $\Delta m^2 \geq 10~{\rm eV}^2$, one excludes oscillations with mixing
angles $\sin^22\theta_{\mu e} > 2\times 10^{-3}$ at 90\% C.L.  Finally,
because the $\nu_\mu$ beam at CERN has about a 1\% $\nu_e$ admixture, both
experiments are also able to exclude $\nu_e\to\nu_\tau$ oscillations for
$\Delta m^2\geq 10~{\rm eV}^2$, but now only for mixing angles
$\sin^22\theta_{e\tau} \geq 10^{-1}$, at 90\% C.L.

\begin{figure}
\begin{center}
\epsfig{file=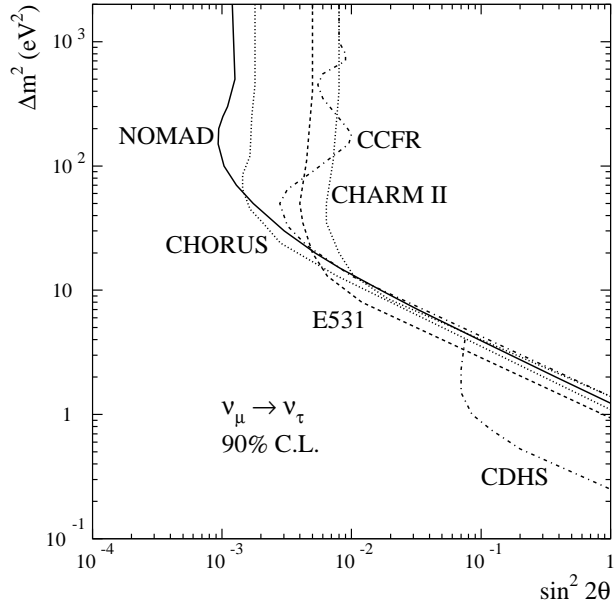,height=3.5in}
\caption[]{Bounds on {$\nu_\mu\to \nu_\tau $} oscillations for large {$\Delta m^2$},  from [27].}
\end{center}
\end{figure}

The situation regarding neutrino oscillations is much less clear cut in the
region $\Delta m^2 \leq 10~{\rm eV}^2$.  Here there are limits from
past accelerator and reactor searches for oscillations and 
recent bounds from the
KARMEN experiment. \cite{karmen}. However, there is also some evidence for $\nu_\mu\to\nu_e$
oscillations coming from the LSND experiment.\cite{LSND}  Let me begin by briefly
detailing the results of this last experiment first.  LSND studies neutrinos originating from pions
produced at rest at the LAMPF beam stop by an 800 MeV proton beam.  These
neutrinos, which are produced in the chain $\pi^+\to\mu^+\nu_\mu\to e^+\nu_e\bar\nu_\mu\nu_\mu$, have average momenta in the 30 to 50 MeV range.
What LSND looks for is the oscillation of the $\bar\nu_\mu$ produced in
$\mu^+$ decay into a $\bar\nu_e$, using a delayed coincidence in a target
30 meters from the beam dump.  If $\bar\nu_\mu\to\bar\nu_e$ oscillations take
place, the $\bar\nu_e$ inverse $\beta$-decay in the target ($\bar\nu_ep\to
e^+n$) produces a prompt photon from $e^+e^-$ annihilation, while 
the produced neutron gives a delayed photon, as a result of the process $np\to d\gamma$.

The LSND experiment observes an excess of $e^+\gamma$ coincidence events which, if
interpreted as  $\bar\nu_\mu\to\bar\nu_e$ oscillations, give a
substantial {\bf allowed region} in the $\Delta m^2-\sin^22\theta$ plane.
However, as I mentioned above, other experiments performed in the past,\cite{past} as well as the
recent KARMEN experiment,\cite{karmen} exclude almost all of this allowed
region.  Furthermore, new data from the KARMEN2 detector\cite{K2} which became available in summer 1998 appeared to
exclude even the small remaining allowed region for LSND!

This rather confusing situation is displayed in Fig. 4. It was discussed in some detail in the
summary talk of Janet Conrad at the 1998 Vancouver International Conference
on High Energy Physics.\cite{Conrad}  As one can see from Fig. 4, a
combination of the BNL 776 data and the Bugey reactor data only leaves the region
between $0.2~{\rm eV}^2 < \Delta m^2 < 4~{\rm eV}^2$ as an ``allowed" region
for the LSND signal.  However, this region is essentially excluded by the KARMEN2 
data, if one uses the 90\% C.L. bound from this experiment.  However, this
result itself is somewhat anomalous, since the 90\% C.L. 
sensitivity for
KARMEN2 is actually below the LSND signal.\footnote{The 90\% C.L. for KARMEN2
of Fig. 4 uses data only from the initial part of their run-where no
background events were seen, even though 3 events were expected.  Additional
data from KARMEN2 now appears to have the number of background events expected.\cite{WIN99}
As a result, it looks like the full KARMEN2 results
will probably be closer to the 90\% C.L. sensitivity line in Fig. 4.}
Furthermore, the LSND experiment has also looked for $\nu_\mu\to\nu_e$
oscillations by studying $\nu_e$ quasielastic scattering events and the collaboration, again,
find an excess of events.\ cite{LSND2}  If interpreted as resulting from oscillations,
this additional signal gives a $\Delta m^2-\sin^22\theta$ allowed
region  which is consistent with that obtained by the $\bar\nu_\mu\to\bar\nu_e$ analysis.

It is difficult to make strong statements at this stage.  The best that one
can say is that there are hints of $\nu_\mu\to\nu_e$ oscillations in the
region $0.2~{\rm eV}^2 < \Delta m^2 < 4~{\rm eV}^2$, with rather small mixing
angles $\sin^22\theta_{\mu e} \sim 10^{-2}$.

\begin{figure}
\begin{center}
\epsfig{file=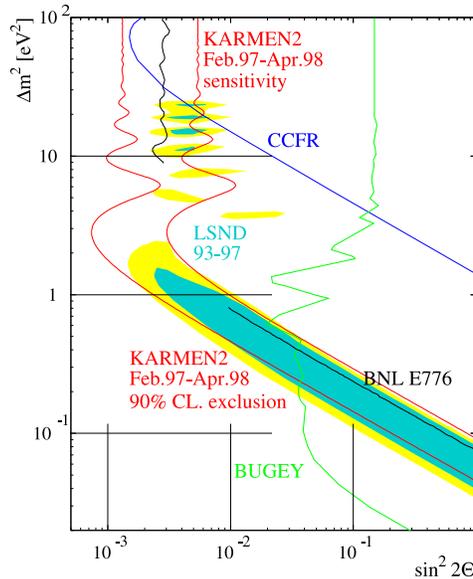,height=3in}
\caption[]{Summary of the experimental situation for {$\bar\nu_\mu\to\bar\nu_e$}  oscillations at the time of the Vancouver Conference, from Ref. [28].}
\end{center}
\end{figure}

\subsection{Atmospheric Neutrino Oscillations.}

Large underground detectors, originally conceived to search for proton decay,
are sensitive to the flux of neutrinos produced in the atmosphere.  These
neutrinos are mostly produced through the decay of pions, with the
$\pi\to\mu\to e$ chain producing two $\nu_\mu$ neutrinos and antineutrinos
for each $\nu_e$ neutrino and antineutrino.  One has known since the early
1990's that the observed flux of $\nu_\mu$'s appeared to be much smaller than
expected, with the ratio\cite{ratio}
\begin{equation}
R = \frac{\left(\frac{\nu_\mu}{\nu_e}\right)_{\rm observed}}
{\left(\frac{\nu_\mu}{\nu_e}\right)_{\rm expected}} \simeq 0.6~.
\end{equation}

Although the anomalous ratio $R$ could be the result of neutrino oscillations,
strong evidence for neutrino oscillations only emerged in summer 1998 from
the SuperKamiokande experiment.  The SuperKamiokande collaboration\cite{superK}
reported a pronounced zenith angle dependence for the flux of multi-GeV
$\nu_\mu$ neutrinos, but no such dependence from $\nu_e$ neutrinos.  For neutrino energies in the
multi-GeV range, the neutrino fluxes are not affected by geomagnetic
effects in an asymmetric fashion. Thus one expects 
the observed neutrino signal to be {\bf up-down symmetric}.  As can be seen in Fig. 5, the
SuperKamiokande data for multi-GeV $\nu_\mu$'s is clearly up-down asymmetric.
There are 139 up-going $\nu_\mu$ compared to 256 down going events.  The
observed asymmetry
\begin{equation}
\left(\frac{U-D}{U+D}\right)^{\rm Multi-GeV}_{\nu_\mu} =
-0.296\pm 0.048 \pm 0.010
\end{equation}
is a $6\sigma$ effect.  The corresponding asymmetry for multi-GeV $\nu_e$
\begin{equation}
\left(\frac{U-D}{U+D}\right)^{\rm Multi-GeV}_{\nu_e} =
-0.036\pm 0.067 \pm 0.020
\end{equation}
is quite consistent with zero.

\begin{figure}
\begin{center}
\epsfig{file=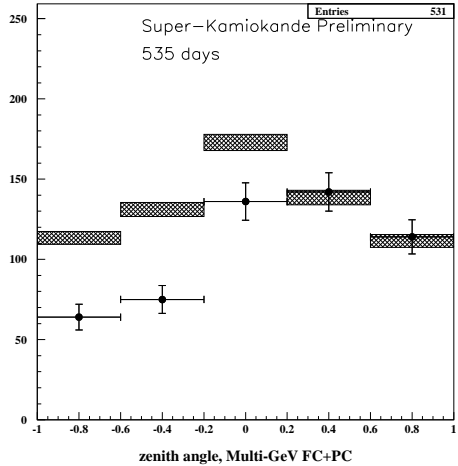,width=3in}
\caption{SuperKamiokande results on multi-GeV {$\nu_{\mu}$} events.}
\end{center}
\end{figure}

The SuperKamiokande collaboration\cite{superK} interprets these results as
evidence for $\nu_\mu\to\nu_X$ oscillations, with $\nu_X$ some other
neutrino species.  This is most dramatically demonstrated in Fig. 6 where the
ratio of data to Monte Carlo is plotted as a function of $L/E_\nu$ for both
$\nu_e$ and $\nu_\mu$ events.  No $L/E_\nu$ dependence is seen in the
$\nu_e$ data, but the $\nu_\mu$ data drops down to a value of 1/2 for
$L/E_\nu \geq 10^3~{\rm Km/GeV}$.  Recalling the simple 2-neutrino formula
for the probability of oscillations [Eq. (78)], Fig. 6 suggest immediately
that $\Delta m^2\sim 10^{-3}~{\rm eV}^2$ and that the mixing angle
$\theta$ is near maximal.  This is confirmed by a more detailed analysis,
which for $\nu_\mu\to\nu_X$ oscillations gives $\sin^22\theta=1$ and
$\Delta m^2= 2.2\times 10^{-3}~{\rm eV}^2$ as the best fit point.

\begin{figure}
\begin{center}
\epsfig{file=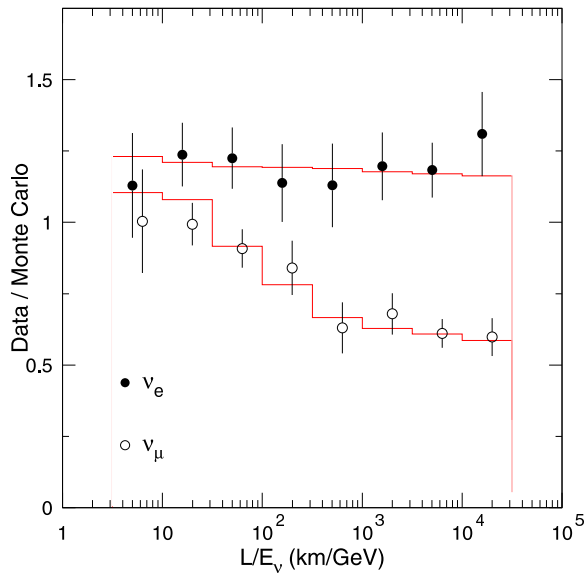,width=3in}
\caption{ Plot of neutrino signals in the SuperKamiokande experiment as a function of {$L/E_\nu$}.}
\end{center}
\end{figure}

It is unlikely, however, 
that the SuperKamiokande results are due to $\nu_\mu\to\nu_e$
oscillations (that is, that $\nu_X\equiv \nu_e$).  First, the region in
the $\Delta m^2-\sin^22\theta$ plane favored by the SuperKamiokande results, is
almost totally excluded already by the null results of the CH00Z reactor
experiment\cite{CH00Z} which looks at $\nu_e$ oscillations into another
neutrino species.  Furthermore, the up-down ratio (129) for the $\nu_e$
flux is more than $3\sigma$ away from what one would expect if one were
dealing with $\nu_\mu\to\nu_e$ oscillations, where one expects
\begin{equation}
\left(\frac{U-D}{U+D}\right)^{\rm theory}_{\nu_\mu\to\nu_e} =
0.205~.
\end{equation}
If there are only three neutrino species, then most likely what is being seen
in SuperKamiokande are $\nu_\mu\to\nu_\tau$ oscillations.  However, at this
stage, it is not possible to rule out the possibility that $\nu_X$ may be a
sterile neutrino $\nu_s$.\footnote{Sterile neutrinos are, by definition,
$SU(2)\times U(1)$ singlets.  Because they do not 
couple to the $Z$, they are not excluded by the neutrino counting results from LEP.}

The SuperKamiokande results\cite{superK} are consistent with previous
Kamiokande results,\cite{Kam} which also had indicated a (less pronounced)
zenith angle dependence of the $\nu_\mu$ flux.  Although the 
$\Delta m^2-\sin^22\theta$ regions for SuperKamiokande and Kamiokande 
do not appear to overlap much, the 90\% C.L. region of SuperKamiokande is
``more-significant", since the Kamiokande best fit has 
$\sin^22\theta = 1.35$.
In fact, recent results presented by SuperKamiokande at DPF 99,\cite{DPF99}
with more data collected, 
help span the gap, indicating even more clearly the
consistency of all data with each other.

In addition, data from other underground experiments (Soudan\cite{Soudan} and
MACRO\cite{macro}) as well as other phenomena---like the flux of upward going
muons \cite{Kam2} produced by $\nu_\mu$ interactions in the
earth---when interpreted in a neutrino oscillation framework, are totally
consistent with the SuperKamiokande $\nu_\mu$ zenith angle results.  Fig. 7
summarizes all this information in one graph.  This figure provides quite
strong evidence in favor of neutrino masses and is probably the strongest
evidence we have to date for physics beyond the Standard Model.

\begin{figure}
\begin{center}
\epsfig{file=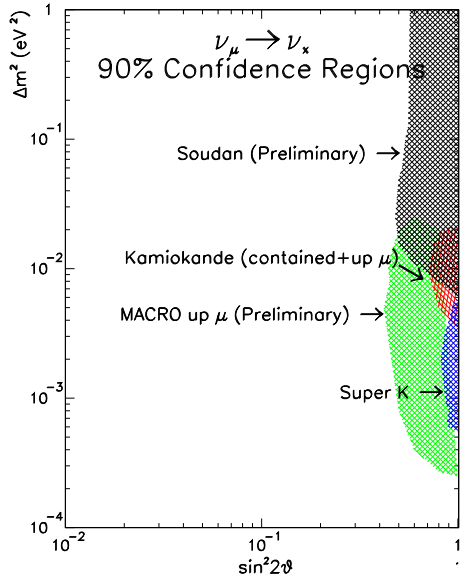,height=3in}
\caption{Evidence for atmospheric neutrino oscillations from all experiments, from Ref. [24].}
\end{center}
\end{figure}

\subsection{Solar Neutrinos}

The study of the solar neutrino flux was started in the early 1970's by Ray Davis
and his group.\cite{Davis}  At present there are five different experiments which
give information on solar neutrinos (Homestake,\cite{H} Gallex,\cite{Gallex}
SAGE,\cite{Sage} Kamiokande\cite{Kamioka} and SuperKamiokande\cite{SuperK2}) and
all five have some bearing on the issue of neutrino oscillations.  In fact, roughly speaking,
all five experiments see approximately half of the expected rate, as shown
in Fig. 8.  However, these experiments are sensitive to different parts of
the solar neutrino spectrum, because the reactions they use to detect solar
neutrinos in their detectors have different thresholds.  SAGE and Gallex
study the reaction $\nu_e + {}^{71}{\rm Ga}\to {}^{71}{\rm Ge} + e^-$, which
has a threshold of 0.23 MeV.  Homestake looks for the excitation of
chlorine ($\nu_e + {}^{37}{\rm Cl}\to {}^{37}{\rm Ar} + e^-$) which has a
0.8 MeV threshold.  The water Cerenkov detectors, Kamiokande and
SuperKamiokande, study elastic $\nu_ee$  scattering and their threshold is in
the neighborhood of 6.5 MeV.\footnote{SuperKamiokande is making strong efforts
to move this threshold down to 5.5 MeV.}

\begin{figure}
\begin{center}
\epsfig{file=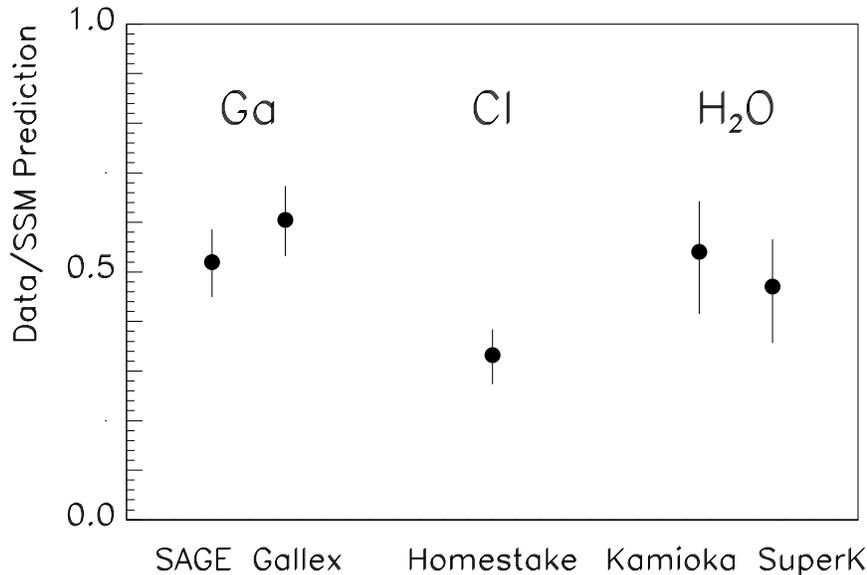,height=3in}
\caption{Rates seen by the diferent solar neutrino experiments, compared to the expectations of the Standard Solar Model. [49]}
\end{center}
\end{figure}

It has long been felt that the observed discrepancy between the neutrino signals detected and the expectations of the, so called, Standard Solar Model(SSM) \cite {BP} is not due to defects in this model but to the presence of some new physical phenomena. One of the principal arguments in favor of this latter solution to the solar neutrino puzzle, has to do with the details of the signal expected in each experiment. Because of the quite different threshold involved, each of the solar neutrino experiments, in fact, feels different pieces of the neutrino producing reactions in the
solar cycle.  For example, the Gallium experiments are the only ones which are
sensitive to neutrinos originating in the $pp$ cycle (the main solar cycle),
with these neutrinos contributing about 50\% of the expected rate.  The
Homestate detector mostly measures neutrinos from ${}^8B$, although it is
also sensitive to ${}^7{\rm Be}$ neutrinos.  Finally, because of their high
threshold, the big water Cerenkov 
detectors only see Boron neutrinos.  These 
circumstances make it difficult to argue for an astrophysical solution to the
solar neutrino deficit.  Much more natural is to imagine that this deficit
arises as a result of neutrino oscillations.

There are two distinct neutrino oscillation solutions to the solar
neutrino problem.  Because roughly all experiments are reduced by about a
factor of two from expectations, it is possible to fit the data by using
vacuum neutrino oscillations $\nu_e\to\nu_X$.  Clearly, for this fit one
must appeal to large mixing angles and assume a tiny $\Delta m^2$.  Since
$E_{\nu}\sim {\rm MeV}$ and the earth-sun distance $L\sim 10^{11}~{\rm m}$,
typically $\Delta m^2\sim 10^{-11}~{\rm eV}^2$.  However, because the 
Homestake result is only about 30\% of the predicted value, one has to fine-tune
the parameters, so that only a few ``just so" regions are favored. \cite{Barger} A recent
``just so" fit by Bahcall, Krastev, and Smirnov\cite{BKS} is shown in
Fig. 9.  

\begin{figure}
\begin{center}
\epsfig{file=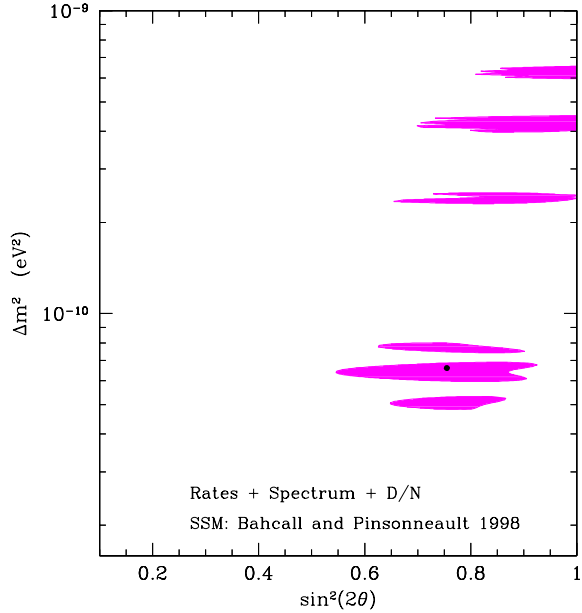,height=3.5in}
\caption[]{"Just so" solar neutrino fit, from [51].}
\end{center}
\end{figure}

In my opinion, much more interesting that the above
``solution" is the possibility that the solar neutrino results are a
reflection of matter induced oscillations (the MSW effect we discuss in
Section V).  In the sun, the electron density to a good approximation, can be
characterized by an exponential profile function \cite{Bah}
\begin{equation}
N_e(r) = N_e(0) e^{-\frac{10r}{R_0}}~.
\end{equation}
The central density $N_e(0) \simeq 10^{26}~{\rm cm}^{-3} \simeq 10^{12}~
{\rm (eV)}^3$ is rather high and for appropriate values of $\Delta m^2$ and
$\sin^22\theta$ can exceed the critical MSW density.  For instance, for
$\theta$ small, $\Delta m^2\simeq 10^{-5}~{\rm eV}^2$ and $|p|\sim 3~{\rm MeV}$
\begin{equation}
N_e^{\rm crit} = \frac{\Delta m^2\cos^2\theta}{\sqrt{2}|p|G_F} \sim
10^{11}~{\rm (eV)}^3~.
\end{equation}
It follows from Eq. (132) that, for these parameters, $\nu_e$'s produced in the
core of the sum (where $N_e(o) \gg N_e^{\rm crit}$) as they radiate outward
go through a region with $N_e\sim N_e^{\rm crit}$ and can oscillate to
$\nu_\mu$'s (or other neutrino types) without paying a mixing angle penalty,
since $\left.\sin2\theta_M\right|_{N_e^{\rm crit}}\to 1$.

The actual calculation of  what happens in the sun is rather complicated, \cite{Petcov} since the
density $N_e$ changes along the neutrino trajectory.  Since $N_e = N_e(t)$,
the matter Hamiltonian of Eq. (94) is now {\bf time dependent}:
\begin{equation}
H_{\rm matter}(t) = \frac{\Delta m^2}{4|p|} \sin2\theta
\sigma_1 - \left(\frac{\Delta m^2}{4|p|}\cos2\theta - \frac{G_F}{\sqrt{2}}
N_e(t)\right) \sigma_3~.
\end{equation}
Although one can diagonalize this Hamiltonian, the resulting mixing angles
and energies will be time dependent:
\begin{equation}
E_{1,2}^M(t) = \pm\left[\left(\frac{\Delta m^2}{4|p|} \cos2\theta -
\frac{G_FN_e(t)}{\sqrt{2}}\right)^2 + \left(\frac{\Delta m^2}{4|p|}
\sin2\theta\right)^2\right]^{1/2}
\end{equation}
\begin{equation}
\tan 2\theta_M(t) = \frac{\frac{\Delta m^2}{4|p|}\sin2\theta}
{\frac{\Delta m^2}{4|p|}\cos2\theta - \frac{G_FN_e(t)}{\sqrt{2}}}~.
\end{equation}
Because of this time dependence, it is no longer true that the
states $|\nu_1^M(t)\rangle$ and $|\nu_2^M(t)\rangle$ are actual eigenstates.
In fact, transitions can occur between these states.
A simple calculation \cite {Petcov} shows that the states $|\nu_i^M(t)\rangle$ obey a
coupled Schr\"odinger equation
\begin{equation}
i\frac{\partial}{\partial t}\left[
\begin{array}{c}
|\nu_1^M(t)\rangle \\ |\nu_2^M(t)\rangle
\end{array} \right] = \left[
\begin{array}{cc}
E_1^M(t) & i\frac{\partial}{\partial t}\theta_M(t) \\
i\frac{\partial}{\partial t}\theta_M(t) & E_2^M(t)
\end{array} \right] \left[
\begin{array}{c}
|\nu_1^M(t)\rangle \\ |\nu_2^M(t)\rangle
\end{array} \right]~.
\end{equation}
If
\begin{equation}
|E_2^M(t)-E_1^M(t)| \gg 
\left|2\frac{\partial}{\partial t} \theta_M(t)\right|
\end{equation}
then transiting between $|\nu_1^M(t)\rangle$ and $|\nu_2^M(t)\rangle$
will be relatively unimportant and one has an {\bf adiabatic} situation.
For an exponential density profile, Eq. (137) is satisfied at $N_e^{\rm crit}$ 
provided that\cite{Petcov}
\begin{equation}
\frac{\Delta m^2\sin^22\theta}{2\cos2\theta} \gg 2\times 10^{-8}~
{\rm (eV)}^2~.
\end{equation}

For the adiabatic case, one can use our discussion of matter oscillations to give a qualitative picture of how the MSW mechanism could work in the sun.  Because at the solar
core $N_e(o) \gg N_e^{\rm crit}$, according to Eq. (106)
$|\nu_e\rangle \simeq |\nu_2^M(o)\rangle$.  Because we are assuming
adiabaticity, as the neutrinos diffuse out of the core 
of the sun, the state
$|\nu_2^M(o)\rangle$ will evolve into $|\nu_2^M(t)\rangle$.  That is, there
are no transitions in the sun.  Thus, when the neutrinos exit the sun, the state
$|\nu_2^M(t_{\rm surface})\rangle$ will just simply become $|\nu_2\rangle$.
Because
\begin{equation}
\langle\nu_e|\nu_2\rangle = \sin\theta~,
\end{equation}
it follows that, in this case,
\begin{equation}
P^{\rm adiabatic}_{\rm solar} (\nu_e\to\nu_e;L) = \sin^2\theta~.
\end{equation}

A more careful analysis shows that there are actually two MSW solutions, one adiabatic and one non-adiabatic,\cite{solarsol} 
both having $\Delta m^2\sim 10^{-5}~{\rm eV}^2$.  The
adiabatic solution has large mixing angles $\sin^22\theta\simeq 1$.  Hence,
according to Eq. (140), $P^{\rm adiabatic}_{\rm solar}(\nu_e\to\nu_e;L)
\simeq 1/2$ so, indeed, roughly half the flux is lost.  The non-adiabatic
solution has $\sin^22\theta\sim 5\times 10^{-3}$.  Furthermore, a rather
large range in the $\Delta m^2-\sin^22\theta$ plane is eliminated by the
absence of a day/night effect, which would be a sign of matter oscillations
in the earth.  The favored MSW regions are depicted in Fig. 10.

\begin{figure}
\begin{center}
\epsfig{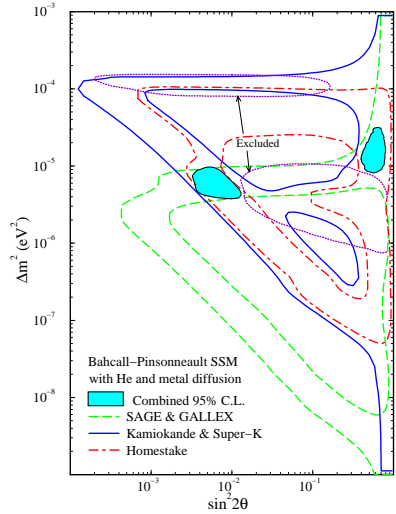}
\caption{Regions in the the {$\Delta m^2-\sin^22\theta$} plane favored by the MSW explanations of the solar neutrino data, from Ref. [51].}
\end{center}
\end{figure}

I want to close this brief discussion of solar neutrinos, and in
particular of the MSW explanation of the solar data, by making a more 
quantitative remark.  To fit the solar neutrino data using the MSW effect
requires that the probability $P(\nu_e\to\nu_e;L)$ have considerable energy
dependence.  The required energy dependence is shown in Fig. 11. I
indicate also in this figure at what energies the neutrinos produced in the various solar
reactions are effective.  One sees from Fig. 11 that essentially all
$pp$ neutrinos survive, the Berylium neutrinos disappear and the flux of Boron neutrinos is roughly halved.  This reconciles nicely with what is seen in the
data, as detailed in Table 3.

\begin{figure}
\begin{center}
\epsfig{file=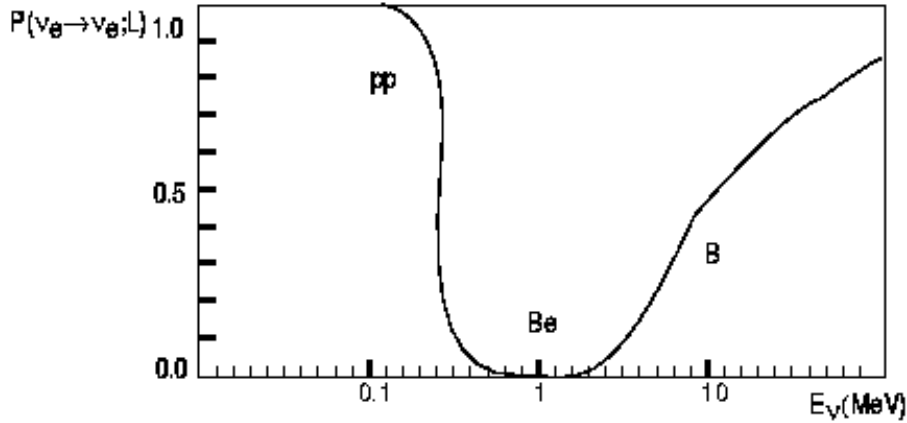,height=2.5in}
\caption{Energy dependence needed for  {$P(\nu_e\to\nu_e;L)$} to fit the solar neutrino data.[54]}
\end{center}
\end{figure}

\begin{center}
{{\bf Table 3.}~~Summary of neutrino flux observations and expectations  
of solar \\ neutrino experiments}
\end{center}
%\begin{tabular}{l}
%\vskip.3cm
%\hline
\begin{center}
{\begin{quote}
${}^{37}{\rm Cl}~(13\%~{\rm Be};~80\%~{\rm B})$ \\ 
$(2.56\pm 0.23)~{\rm SNU}~ {\rm Ref.} [44]~~~\left[\mbox{predicted SSM}:~
\left(7.7^{+1.2}_{-1.0}\right)~{\rm SNU}\right]$ \\
\end{quote}
\begin{quote}
${}^{71}{\rm Ga}~(51\%~{\rm pp};~ 15\%~{\rm Be};~ 12\%~{\rm B})$ \\ 
$(77.5\pm 7.7)~{\rm SNU}~ {\rm Ref.} [ 45]~~~[\mbox{predicted SSM}:~
(129\pm 8)~{\rm SNU}]$ \\ 
$(66.6\pm 8.0)~{\rm SNU}~{\rm Ref.} [46] $ \\
\end{quote}
\begin{quote}
${\rm Water}~\check C~(100\%~{\rm B})$ \\ 
$2.80\pm 0.38~{\rm SNU}~{\rm Ref}. [47]~~~[\mbox{predicted SSM}:~
\left(5.15^{+1.0}_{-0.7}\right)~{\rm SNU}]$ \\ 
$2.44 \pm 0.10~{\rm SNU}~{\rm Ref.} [48]$ \\
\end{quote}}
\end{center}

\section{Theoretical Implications}

I summarize in Fig. 12 where one is at present on the issue of neutrino
masses and mixings. This figure
collects together all the neutrino oscillation 
evidence, as well as the hints for oscillations, which we have at
the moment.  As is clear from the figure there are three regions 
suggested
in the $\Delta m^2-\sin^22\theta$ plane.  The strongest evidence is that for
atmospheric neutrino oscillations coming from the SuperKamiokande zenith angle
data.  Here the suggested parameters are $(\Delta m^2)\sim 3\times 10^{-3}~
{\rm eV}^2$, $\sin^22\theta\sim 1$ with $\nu_\mu\to\nu_X~(\nu_X\not= \nu_e)$.
Solar neutrinos also are strongly suggestive of oscillations.  Interpreting
the data this way leads to $(\Delta m^2)\sim 10^{-5}~{\rm eV}^2$, with
$\sin^22\theta\sim 1$ or $\sin^22\theta\sim 5\times 10^{-3}$, for
MSW $\nu_e\to\nu_X$ oscillations, or $\Delta m^2\sim 10^{-11}~{\rm eV}^2$
and $\sin^22\theta\sim 1$, for ``just-so" $\nu_e\to\nu_X$ oscillations.
The weakest hint for oscillations probably is that of LSND, because of other
contrary evidence.  At any rate, the suggested region here is $\Delta m^2\sim
5\times 10^{-1}~{\rm eV}^2$ and $\sin^22\theta\sim 10^{-2}$ for
$\nu_e\to\nu_\mu$ oscillations.

\begin{figure}
\begin{center}
\epsfig{file=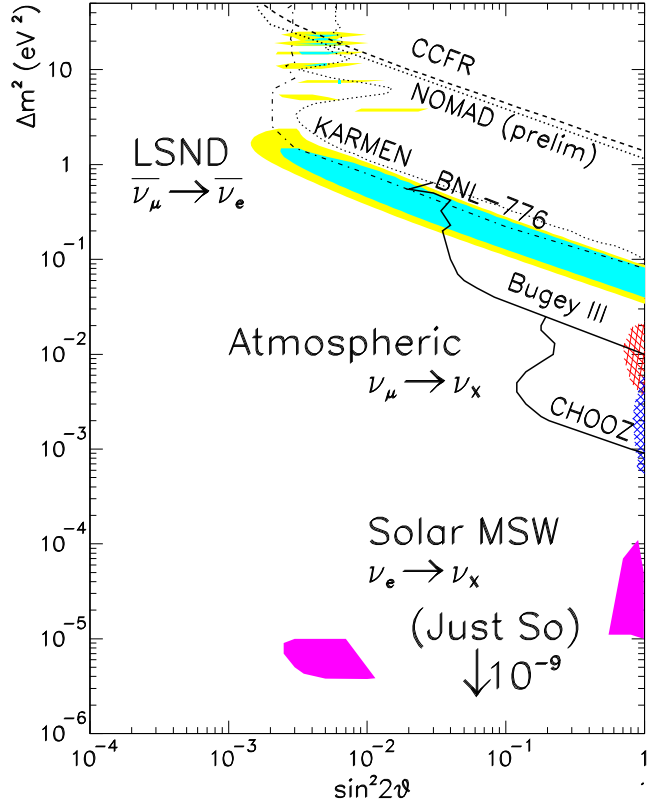,height=4.5in}
\caption{Summary of evidence of, and hints for, neutrino oscillations.[24]}
\end{center}
\end{figure}

Besides neutrino oscillation phenomena, there is really no other direct
evidence for neutrino masses.  However, both $\beta$-decay and neutrinoless
double $\beta$-decay put rather strong bounds on neutrino masses connected to
$\nu_e$.  From $\beta$-decay the largest neutrino mass obeys the bound: ``$m_{\nu_e}$" 
$< 3.9~{\rm eV}$ (90\% C.L.).  Double $\beta$-decay, bounds the
Majorana mass $\langle m_{\nu_e}\rangle_{ee}$ even more strongly:  $\langle
m_{\nu_e}\rangle_{ee} < 0.2~{\rm eV}$ (90\% C.L.).  These bounds are
interesting, since they are close to the kind of neutrino masses which could
have substantial cosmological influence.  Using the central value for the
Hubble parameter [c.f. Eq.(124)], I note that $\sum_i m_{\nu_i} = 30~{\rm eV}$, 6 eV, 2 eV
correspond, respectively, to neutrinos closing the Universe, to neutrinos
being 20\% of the dark matter in the Universe (assuming $\Omega_M\simeq 1$),
and to neutrinos being 20\% of the dark matter, with $\Omega_M\simeq 0.3$.  The last case is, perhaps, the one that is most cosmologically realistic. \cite{cosmo}  This stresses the
the importance of continuing the search for neutrino  masses in the eV
range.

Concentrating only on the SuperKamiokande evidence for neutrino masses
already has important implications.  Taking
\begin{equation}
\Delta m^2 = m_3^2-m_2^2\sim 3\times 10^{-3}~{\rm eV}^2
\end{equation}
gives one already a {\bf lower bound} on 
some neutrino mass:  $m_3\geq 5\times 10^{-2}~{\rm eV}$.  This mass value, in turn, gives a lower bound for the cosmological
contribution of neutrinos
\begin{equation}
\Omega_\nu \geq \frac{m_3}{92~{\rm eV}~h^2} \sim 1.5\times 10^{-3}~.
\end{equation}
Although this number is far from that needed for closure of the Universe, I note that the $\Omega_\nu$ of
Eq. (142) is comparable to the contribution of luminous matter to the energy
density of the Universe\cite{luminous}
\begin{equation}
\Omega_{\rm luminous}\sim (3-7)\times 10^{-3}~.
\end{equation}
So, from SuperKamiokande we learn that the neutrino contribution to the energy density of the Universe is the same as,
in the words of Carl Sagan, that of ``billions and billions of stars"!

For particle physics, a value of $m_3\sim 5\times 10^{-2}~{\rm eV}$ is also
quite interesting.  If we use either the simple see-saw formula of Eq. (55)
or the GUT relation (66), the identification
\begin{equation}
m_3 = \left\{
\begin{array}{l}
\frac{[(m_D)_3]^2}{m_S} \sim \frac{m_t^2}{m_S} \\
m_T \simeq \frac{\langle\phi^0\rangle^2}{\Lambda}
\end{array} \right.
\end{equation}
give comparable values for $m_S$ and $\Lambda$:
\begin{equation}
m_S \sim \Lambda\sim 10^{15}~{\rm GeV}~.
\end{equation}
Of course, these values are only justifiable in specific models where one
has a bit more control of other constants, which are taken above 
all to be of $O(1)$.

If one goes beyond the SuperKamiokande data, then many
theoretical scenarios emerge.  Unfortunately, in general, these scenarios
mostly reflect the prejudices one has regarding the data.  Nevertheless,
it is useful to briefly discuss two differing broad theory scenario.  In
the first scenario, one assumes that all hints for oscillations seen are
true.  In the second, one disregards some oscillation hints.  In most cases,
the discarded data is that of LSND.

If one believes all hints for neutrino oscillations, since there are three
different $\Delta m^2$ involved, the neutrino mass matrix $M$ necessarily is a
$4\times 4$ matrix.\footnote{There have been attempts to ``stretch" some of
the data, so that all hints can be accounted for with only two different
$\Delta m^2$.  These attempts\cite{stretch} seem rather forced to me.}
To get a $4\times 4$ neutrino matrix one adds to the usual three neutrinos a
sterile neutrino $\nu_s$.  Most 4 neutrino models attempt to fit all data, 
since this was after all the reason for introducing the fourth neutrino.
The most promising scenario\cite{fitall} has two pairs of quasi-Dirac 
neutrinos split by a small mass difference.\footnote{A quasi-Dirac neutrino
pair reduces to a Dirac neutrino, as the mass difference between the pair
vanishes.}  The heaviest pair ($m_2$ and $m_3$) have masses of order
0.5 eV and contribute $\Omega_\nu\simeq 0.03$ to cosmology.  The atmospheric
neutrino oscillations involve this pair, so that $\Delta m^2_{23}\sim 3\times
10^{-3}~{\rm eV}^2$.  The second pair ($m_1$ and $m_4$) are much lighter, with
mass around $10^{-2}-10^{-1}~{\rm eV}$.  Their mass difference
$\Delta m^2_{14}\sim 10^{-5}~{\rm eV}^2$ is what enters in solar neutrino
oscillations.  The LSND result is explained as an oscillation between the
light pair and the heavy pair, with $\Delta m^2\sim m^2_3\sim 0.6~{\rm eV}^2$.  In this scheme, the solar neutrino oscillations involve 
oscillations of $\nu_e$
to a sterile neutrino, while the atmospheric neutrino 
oscillation is $\nu_\mu\to\nu_\tau$ and LSND $\nu_e\to\nu_\mu$.  Although
this scheme works phenomenologically, theoretically it is difficult to get
light sterile neutrinos almost degenerate with ordinary neutrinos.

Different patterns arise if one is prepared to disregard some of the
neutrino oscillation limits.  If one disregards, in particular, the results 
from LSND then the CH00Z bound,\cite{CH00Z} and the quite different mass
squared differences involved in atmospheric and solar neutrino oscillations,
suggest a very simple 3-neutrino mixing matrix.  CH00Z suggests that
$\theta_{13}\simeq 0^o$.  On the other hand, atmospheric neutrino oscillations
suggest $\theta_{23}\simeq 45^o$.  Finally, depending on what solar neutrino
oscillation solution one picks, the angle $\theta_{12}$ can either be large or
small.  Thus, neglecting possible CP violating phases, the neutrino mixing
matrix looks like\cite{Alta}
\begin{eqnarray}
U &\simeq \left[
\begin{array}{ccc}
1 & 0 & 0 \\ 0 & \frac{1}{\sqrt{2}} & -\frac{1}{\sqrt{2}} \\
0 & \frac{1}{\sqrt{2}} & \frac{1}{\sqrt{2}} 
\end{array} \right] \left[
\begin{array}{ccc}
1 & 0 & 0 \\ 0 & 1 & 0 \\ 0 & 0 & 1
\end{array} \right] \left[
\begin{array}{ccc}
c_{12} & -s_{12} & 0 \\ s_{12} & c_{12} & 0 \\ 0 & 0 & 1
\end{array} \right] \\ \nonumber
&= \left[
\begin{array}{ccc}
c_{12} & -s_{12} & 0 \\
\frac{s_{12}}{\sqrt{2}} & \frac{c_{12}}{\sqrt{2}} & -\frac{1}{\sqrt{2}} \\
\frac{s_{12}}{\sqrt{2}} & \frac{c_{12}}{\sqrt{2}} & \frac{1}{\sqrt{2}}
\end{array} \right]~,
\end{eqnarray}
where $c_{12} = \cos\theta_{12};~s_{12}=\sin\theta_{12}$.  However, even
with $U$ of the  above form, there are many open questions to answer.
For instance, is maximal mixing $(\theta_{12}\simeq 45^o)$ allowed?  Are
nearly degenerate neutrino masses $(m_1\simeq m_2\simeq m_3\simeq 
0.5~{\rm eV})$ allowed?  What neutrino mass matrix gives rise to this
particular mixing matrix?

These questions cannot really be answered in a straightforwad manner, without
making some more assumptions.  There is really a lot of freedom.
Given $U$ and some assumptions for the neutrino mass spectrum $\{m_i\}$
then one can deduce a neutrino mass matrix $M$.  However, recall from our
discussion of the see-saw mechanism, that $M$ itself depends on both the
neutrino Dirac mass $m_D$ and on the right-handed neutrino mass matrix
$m_S$ [cf. Eq. (55)].  Thus, to make progress, even ``knowing" $M$ one has to make 
some assumptions on $m_D$ (or $m_S$) to learn something further.  For instance,
one could use GUTs which naturally ties the matrix $m_D$ in Eq. (55) to the
$u$-quark mass matrix. \cite{Babu}

Some authors have preferred to focus on some simple structure for the
$3\times 3$ matrix $M$. \cite{Barbieri} Two of these are particularly appealing.  
The first of these
has a total degeneracy for the neutrinos, the other is Dirac-like with
an additional massless neutrino.  In the first pattern
\begin{equation}
M = m\left(
\begin{array}{ccc}
1 & 0 & 0 \\ 0 & 0 & 1 \\ 0 & 1 & 0
\end{array} \right)~.
\end{equation}
In this case, $\sum_i m_{\nu_i} = 3m$, so that cosmology impose a bound on $m$,
depending on what one believes $\Omega_\nu$ is.  However, in the degenerate case, one has also that
\begin{equation}
\langle m_{\nu_e}\rangle_{ee} = \sum_i U^2_{ei} m_{\nu_i} = m~.
\end{equation}
The double $\beta$-decay bound then tell us 
that $m < 0.2$ eV. 
If we push $m$ to its upper bound, however, it is difficult to see what
perturbation can then give $\Delta m^2_{\rm atmos}\simeq 3\times 10^{-3}~{\rm eV}^2;~
\Delta m^2_{\rm solar} \sim 10^{-5}~{\rm eV}^2$.

The second simple pattern for neutrino masses has \cite{Hall}
\begin{equation}
M = m\left(
\begin{array}{ccc}
0 & 1 & 1 \\ 1 & 0 & 0 \\ 1 & 0 & 0
\end{array} \right)
\end{equation}
which has a degenerate pair and a zero eigenvalue.  Note that this pattern
conserves $L_e-L_\mu-L_\tau$.  Since for this mass matrix $\Delta m^2_{\rm atmos} = m^2$,
it follows that $m\sim 5\times 10^{-2}~{\rm eV}$.  So in this case,
neutrinos do not contribute much to the energy density of the Universe
$(\Omega_\nu \simeq 3\times 10^{-3})$.  To get solar neutrino oscillations
one has to introduce some perturbation on the mass matrix (149) that
will split the massive degenerate states and give $\Delta m^2_{12} \sim
10^{-5}~{\rm eV}^2$.

\section{Future Experiments}

It seems pretty clear that progress in understanding what is going on in the
neutrino sector can only come from further data.  Fortunately, new data will
be forthcoming in all the relevant $\Delta m^2$ regions.  I want to end
these lectures by briefly discussing these future experiments.

\subsection{Solar Neutrinos}

SuperKamiokande will continue to take data in years to come, thus refining
their present measurements of solar neutrinos.  Furthermore, a real effort is taking place to lower
the neutrino energy threshold further so as to be able to study the shape
dependence of the signal as a function of $E_\nu$.  In addition to this
continuing effort, relatively soon two other experiments will be coming on
line which have considerable promise.  The first of these is SNO (the
Sudbury Neutrino Observatory \cite{SNO}) which uses a Kiloton of D$_2$O.  The
advantage of having heavy water is that it allows 
SNO to study simultaneously both
charged current and neutral current processes.  The charged current process
\begin{equation}
\nu_e + d\to e^- + p+p~,
\end{equation}
like all charged current processes, is sensitive to whether oscillations
have occurred or not.  The neutral current disintegration of the deuteron,
on the other hand, is insensitive to oscillations since it is the same for all
neutrino species $\nu_X$:\footnote{This is only true for  
neutrinos whose neutral couplings to the $Z$ are universal.  It does not
apply to sterile neutrinos.}
\begin{equation}
\nu_X + d\to\nu_X+p+n~.
\end{equation}
Comparison of the rates for the two neutrino 
reactions(150) and (151) should help rule
out possible astrophysical explanations for the solar neutrino puzzle.  The SNO
detector should begin taking data in 1999.

The second solar neutrino experiment of interest is Borexino. \cite{Borexino}  This experiment
is presently under construction at the Gran Sasso Laboratory and should be
ready for data taking in 2001.  Borexino uses 300 tons of scintillator,
which has a relatively low threshold $(E_{\rm thr} > 340~{\rm KeV})$.  As
a result, Borexino should be particularly sensitive to the $E_\nu = 862~{\rm GeV}$ neutrino line 
coming from ${}^7$Be.  Recalling Fig. 11, one sees that if the MSW
explanation is correct, the solar neutrino signal in Borexino should be
significantly below the theoretical expectations.  Indeed, if there are no solar
oscillations, Borexino is supposed to detect about 50 events/day, 
while if the
MSW explanation is true, this number should go down to about 10 events/day.

\subsection{Atmospheric Neutrinos}

Here again SuperKamiokande will continue to integrate data with time.  However,
the $\Delta m^2$ region, will also be probed more directly by using neutrino
beams from accelerators.  Three such long baseline experiments are in different
stages of readiness.  K2K, which uses a neutrino beam from KEK, aimed at
SuperKamiokande 250 Km away, should shortly be operational. \cite{K2K}  MINOS,\cite{MINOS} in the
Soudan Mine, is under construction and will be the target of a dedicated neutrino beam from Fermilab, 730 Km away.  First data should become available
around 2001-2002.  Finally, a variety of proposals exist for experiments in
the Gran Sasso Laboratory, which is 740 Km from CERN, to become targets
of neutrino beams from CERN.

The main advantage that these long baseline experiments have over SuperKamiokande is
that the neutrino beam used is well characterized, both in energy 
and in its
time structure.  Furthermore, these beams also have a higher intensity.
So many possible systematic effects will be under better control.  In the
case of the higher energy Fermilab and CERN beams, it may also be possible to
directly search and detect $\nu_\tau$'s, if these neutrinos are produced in the oscillations.

\subsection{LSND Region}

The $\Delta m^2-\sin^22\theta$ region identified by the LSND experiment as
potentially interesting also  will be explored further.  At Fermilab, there is an
approved experiment, Mini BooNE, \cite{Boone} which will run around 2001-2002, which should
be about a factor of five more sensitive than LSND in a comparable kinematical region.  
With this sensitivity,
it should be quite clear whether $\nu_\mu\to\nu_e$ oscillations with
$\Delta m^2\sim (0.1-1)~{\rm eV}^2$ exist or not.
 
\section*{Acknowledgements}

I am grateful to Juan Carlos D'Olivo and Myriam Mondragon for their wonderful hospitality in Oaxaca. I am also thankful to all the students at the VIII Escuela Mexicana de Particulas y Campos for their  attention and enthusiasm. 
This work was supported in part by the Department of Energy inder contract No. DE-FG03-91ER40662, Task C.

\section*{Appendix A: Dirac and Majorana Masses}
\setcounter{equation}{0}
%\addtocounter{section}{A}

To understand how Dirac and Majorana masses can arise, it is useful to review
here some of the properties of the spinor representations of the Lorentz
group.  The Lorentz group, besides the well known vector and tensor
representations has also spinor representations. It turns out
that there are two inequivalent spinor representations. It is out of
these two-dimensional spinors that one builds up the usual four-dimensional
Dirac spinor $\psi$.

Under a Lorentz transformation, a vector field $V^\mu$ has the well known
transformation
\begin{equation}
V^\mu \to V^{\prime\mu} = \Lambda^\mu_{~\nu} V^\nu~, 
\end{equation}
where the 4-dimensional representation matrices 
$\Lambda$ obey the pseudo-orthogonality
conditions
\begin{equation}
\eta_{\mu\nu} = \Lambda^\alpha_{~\mu} \eta_{\alpha\beta}\Lambda^\beta_{~\nu}~,
\end{equation}
involving the metric tensor
\begin{equation}
\eta_{\mu\nu} = \left[ 
\begin{array}{cccc}
-1 & & & \\ & 1 & & \\ & & 1 & \\ & & & 1
\end{array}
\right]~.
\end{equation}

Besides vector representations, the Lorentz group has two inequivalent
spinor representation.  The corresponding 2-dimensional Weyl spinors are
conventionally denoted by $\xi_a$ and $\dot\xi_a$, known as undotted and
dotted spinors, respectively.  Under Lorentz transformations they transform as
\begin{eqnarray}
\xi_a \to \xi_a^\prime &=& M_a^{~b} \xi_b \\
\dot\xi_a \to \dot\xi_a^\prime &=& M_a^{*b}\dot\xi_b~.
\end{eqnarray}
The $2\times 2$ matrices $M$ and $M^*$, with det $M = {\rm det}~M^*=1$,
provide inequivalent representation of $SL(2,C)$.  Obviously, from the above
it follows that $\dot\xi\sim\xi^*$.

One can establish a relationship between the $2\times 2$ matrices $M$ and
the $4\times 4$ matrices $\Lambda$, since the vector field $V^\mu$ transforms
as $V\sim\xi\otimes\dot\xi$.  For these purposes, it is useful to define a
set of four matrices $\sigma^\mu\equiv (1,\vec\sigma)$, with $\vec\sigma$
being the usual Pauli matrices.  The $2\times 2$ matrix
\begin{equation}
V = \sigma^\mu\eta_{\mu\nu}V^\nu \equiv\sigma^\mu V_\mu
\end{equation}
under a Lorentz transformation transforms as
\begin{equation}
V\to V^\prime = MVM^\dagger = \sigma^\mu V_\mu^\prime~.
\end{equation}
Using Eq. (A1), it follows that
\begin{equation}
\sigma^\mu_{ac}\Lambda_\mu^{~\nu} = M_a^{~b}\sigma_{bd}^\nu M_c^{*d}~.
\end{equation}

Because det $M=1$, the analogue of the scalar product for vectors
$V^\mu\eta_{\mu\nu}V^\nu \equiv V^\mu V_\mu$, for the spinors $\xi$ and
$\dot\xi$ leads to the following Lorentz scalars:
\begin{equation}
\xi_a\epsilon^{ab}\xi_b \equiv \xi_a\xi^b~; ~~~~~
\dot\xi_a\epsilon^{ab}\dot\xi_b \equiv \dot\xi_a\dot\xi^b
\end{equation}
where $\epsilon^{ab} = -\epsilon^{ba}$ and $\epsilon^{12} = 1$.  Similarly,
just as the contraction of a covariant and contravariant metric tensor gives
the identity $[\eta_{\mu\rho}\eta^{\rho\nu} = \delta_\mu^\nu]$, one can
define $2\times 2$ antisymmetric $\epsilon$-matrices, $\epsilon_{ab}$, which
obey
\begin{equation}
\epsilon_{ac}\epsilon^{cb} = \delta_a^b~.
\end{equation}
It follows that $\epsilon_{12} = -1$.

The usual 4-component Dirac spinor $\psi$ is made up of a dotted and an
undotted Weyl spinor:
\begin{equation}
\psi = \left(
\begin{array}{c}
\xi_a \\ \dot\chi^a 
\end{array} \right)
\end{equation}
In this, so called, Weyl-basis the Dirac $\gamma$-matrices $\gamma^\mu$,
which obey the anticommutation relations $\{\gamma^\mu,\gamma^\nu\} = 
-2\eta^{\mu\nu}$, take the form
\begin{equation}
\gamma^\mu = \left(
\begin{array}{cc}
0 & \sigma^\mu \\ \bar\sigma^\mu & 0
\end{array} \right)~.
\end{equation}
Here $\bar\sigma^\mu = (1,-\vec\sigma)$, so that in this basis
\begin{equation}
\gamma^0 = \left(
\begin{array}{cc}
0 & 1 \\ 1 & 0 
\end{array} \right)~; ~~~~~
\gamma^i = \left(
\begin{array}{cc}
0 & \sigma^i \\ -\sigma^i & 0
\end{array} \right)
\end{equation}
and
\begin{equation}
\gamma_5 = i\gamma^0\gamma^1\gamma^2\gamma^3 = \left(
\begin{array}{cc}
-1 & 0 \\ 0 & 1
\end{array} \right)~.
\end{equation}

It follows from the above that $\xi_a$ and $\dot\chi^a$ are {\bf chiral projections} of 
$\psi$:
\begin{eqnarray}
\psi_{\rm L} = \frac{1}{2}(1-\gamma_5)\psi &=& 
\left(
\begin{array}{c}
\xi_a \\ 0
\end{array} \right)~; ~~~~
\overline{\psi_{\rm L}} = \bar\psi \frac{1}{2}(1+\gamma_5) =
(0~~\xi^*_a) \\
\psi_{\rm R} = \frac{1}{2}(1+\gamma_5)\psi &=& \left(
\begin{array}{c}
0 \\ \dot\chi^a
\end{array} \right)~; ~~~~
\overline{\psi_{\rm R}} = \bar\psi \frac{1}{2}(1-\gamma_5) =
(\dot\chi^{a*}~0)~.
\end{eqnarray}
Using these equations, it is easy to see that the Dirac mass term connects
$\xi$ with $\dot\chi$.  Specifically, one has
\begin{equation}
{\cal{L}}_{\rm Dirac} = -m_D(\overline{\psi_{\rm L}}\psi_{\rm R} +
\overline{\psi_{\rm R}}\psi_{\rm L}) = -m_D(\xi^*_a\dot\chi^a +
\dot\chi^{a*}\xi_a)~.
\end{equation}
Recall, however, that dotted spinors are related to the complex conjugate
of an undotted spinor.  Choosing a phase convention where
\begin{equation}
\xi^*_a = \dot\xi_a~; ~~~~
\dot\chi_a^* = \chi_a,
\end{equation}
one can write the Dirac mass term simply as
\begin{equation}
{\cal{L}}_{\rm Dirac} = -m_D(\dot\xi_a\dot\chi^a + \chi^a\xi_a)~.
\end{equation}
In view of Eq. (A9) this term is obviously Lorentz invariant.  However,
Lorentz invariance does not require one to have two distinct Weyl spinors
$\xi$ and $\chi$ to construct a mass term.  Majorana masses, basically, make
use of this ``simpler" option.

One can define a 4-component {\bf Majorana spinor} in terms of the Weyl
spinor $\xi$ and its complex conjugate $\dot\xi$:
\begin{equation}
\psi_M = \left(
\begin{array}{c}
\xi_a \\ \dot\xi^a
\end{array} \right)~.
\end{equation}
Because $\dot\xi^a = \xi^{a*}$, effectively $\psi_M$ has only one independent
helicity projection.  One can choose this projection to be, say,
$(\psi_M)_{\rm L}$:
\begin{equation}
(\psi_M)_{\rm L} = \frac{1}{2}(1-\gamma_5)\psi_M = \left(
\begin{array}{c}
\xi_a \\ 0 
\end{array} \right)~; ~~~~
{\overline{(\psi_M)_{\rm L}}} = \overline{\psi_M} \frac{1}{2}(1+\gamma_5) =
(0~~\dot\xi_a)~.
\end{equation}
One can construct $(\psi_M)_{\rm R}$ by using the charge conjugate matrix
$\tilde C$. In the Weyl basis $\tilde C$ is given by
\begin{equation}
\tilde C = \left[
\begin{array}{cc}
\epsilon_{ab} & 0 \\ 0 & \epsilon^{ab}
\end{array} \right] = \left[
\begin{array}{cccc}
0 & -1 & 0 & 0 \\ 1 & 0 & 0 & 0 \\ 0 & 0 & 0 & 1 \\ 0 & 0 & -1 & 0
\end{array} \right]~.
\end{equation}
Clearly
\begin{eqnarray}
(\psi_M)_{\rm R} &=&
 \left(
\begin{array}{c}
0 \\ \dot\xi^a
\end{array} \right) = \left(
\begin{array}{c}
0 \\ \epsilon^{ab}\dot\xi_b
\end{array} \right) =
\tilde C{\overline{(\psi_M)_{\rm L}}}^T \nonumber \\
\overline{(\psi_M)_{\rm R}} &=& (\xi^a~0) = (\epsilon^{ab}\xi_b~0) =
(\xi_b\epsilon_{ba}~0) = (\psi_M)^T_{\rm L}\tilde C~.
\end{eqnarray}
That is, $(\psi_M)_{\rm R}$ is the charge conjugate of $(\psi_M)_{\rm L}$
(c.f. Eq. (36)):
\begin{equation}
[(\psi_M)_{\rm L}]^c = (\psi_M)_{\rm R}~.
\end{equation}
Because of Eq. (A24) it follows that the Majorana spinor $\psi_M$ obeys a
constraint.  It is {\bf self-conjugate}:
\begin{equation}
\psi_M = \left(
\begin{array}{c}
\xi_a \\ \dot\xi^a
\end{array} \right) =
(\psi_M)_{\rm L} + (\psi_M)_{\rm R} = (\psi_M)_{\rm L} +
[(\psi_M)_{\rm L}]^c~. 
\end{equation}
Hence,
\begin{equation}
\psi_M = [\psi_M]^c~.
\end{equation}

The Majorana mass term
\begin{equation}
{\cal{L}}_{\rm Majorana} = -\frac{1}{2} m_M\overline{\psi_M}\psi_M
\end{equation}
involves a product of $\xi$ with itself and $\dot\xi$ with itself
\begin{equation}
{\cal{L}}_{\rm Majorana} = -\frac{1}{2} m_M\left({\overline{(\psi_M)_{\rm L}}}
(\psi_M)_{\rm R} + {\overline{(\psi_M)_{\rm R}}}
(\psi_M)_{\rm L}\right) = -\frac{1}{2} m_M
(\dot\xi_a\dot\xi^a + \xi^a\xi_a)~.
\end{equation}
Eq. (A28) can also be written purely in terms of $(\psi_M)_{\rm L}$ by using the
charge conjugation matrix $\tilde C$.  Using Eq. (A23) one has also
\begin{equation}
{\cal{L}}_{\rm Majorana} = -\frac{1}{2} m_M\left({\overline{(\psi_M)_{\rm L}}}
\tilde C{\overline{(\psi_M)_{\rm L}}}~^T + (\psi_M)_{\rm L}^T
\tilde C(\psi_M)_{\rm L}\right)~.
\end{equation}
Equally well one can write this mass term entirely as a function of
$(\psi_M)_{\rm R}$.  One finds
\begin{equation}
{\cal{L}}_{\rm Majorana} = -\frac{1}{2}\left((\psi_M)_{\rm R}^T
\tilde C(\psi_M)_{\rm R} + {\overline{(\psi_M)_{\rm R}}}
\tilde C{\overline{(\psi_M)_{\rm R}}}~^T\right)~.
\end{equation}

\end{document}